\documentclass[onecolumn]{aastex6}

\usepackage[T1]{fontenc}  % font encoding
\usepackage{amsmath}      % mathematical notation
\usepackage{mathtools}    % paired delimiters
\usepackage{tensor}       % tensor notation
\usepackage{hyperref}     % links
\hypersetup{colorlinks=true,linkcolor=black,citecolor=black,filecolor=black,urlcolor=black}

\makeatletter
\newcommand{\extref}[1]{\textup{\tagform@{#1}}}
\makeatother

\newcommand{\dd}{\mathrm{d}}
\newcommand{\pp}{\partial}
\newcommand{\dt}{\dd t}
\newcommand{\dr}{\dd r}
\newcommand{\dth}{\dd\theta}
\newcommand{\dph}{\dd\phi}
\newcommand{\dtau}{\dd\tau}
\DeclareMathOperator{\sgn}{sgn}
\DeclarePairedDelimiter{\paren}{\lparen}{\rparen}
\DeclarePairedDelimiter{\abs}{\lvert}{\rvert}
\DeclarePairedDelimiter{\ave}{\langle}{\rangle}
\newcommand{\tens}[2]{\tensor*{#1{\mathstrut}}{#2}}

\newcommand{\msun}{M_\odot}
\newcommand{\pgas}{p_\mathrm{gas}}
\newcommand{\pmag}{p_\mathrm{mag}}

\newcommand{\alphaacc}{\alpha_\mathrm{acc}}
\newcommand{\alphapre}{\alpha_\mathrm{pre}}
\newcommand{\alphaali}{\alpha_\mathrm{ali}}
\newcommand{\alpharey}{\alpha_\mathrm{Rey}}
\newcommand{\alphamax}{\alpha_\mathrm{Max}}

\newcommand{\rphot}{r_\mathrm{ph}}
\newcommand{\uksa}{\tens{u}{_{\mathrm{KS}}^0}}
\newcommand{\uksb}{\tens{u}{_{\mathrm{KS}}^1}}
\newcommand{\uksc}{\tens{u}{_{\mathrm{KS}}^2}}
\newcommand{\uksd}{\tens{u}{_{\mathrm{KS}}^3}}
\newcommand{\ubla}{\tens{u}{_{\mathrm{BL}}^0}}
\newcommand{\ublb}{\tens{u}{_{\mathrm{BL}}^1}}
\newcommand{\ublc}{\tens{u}{_{\mathrm{BL}}^2}}
\newcommand{\ubld}{\tens{u}{_{\mathrm{BL}}^3}}
\newcommand{\ubln}{\tens{u}{_{\mathrm{BL}}^\nu}}
\newcommand{\ublbp}{\tens{u}{_{\mathrm{BL}}^{1'}}}
\newcommand{\ubldp}{\tens{u}{_{\mathrm{BL}}^{3'}}}
\newcommand{\ubblap}{\tens{\bar{u}}{_{\mathrm{BL}}^{0'}}}
\newcommand{\ubblbp}{\tens{\bar{u}}{_{\mathrm{BL}}^{1'}}}
\newcommand{\ubblcp}{\tens{\bar{u}}{_{\mathrm{BL}}^{2'}}}
\newcommand{\ubbldp}{\tens{\bar{u}}{_{\mathrm{BL}}^{3'}}}
\newcommand{\ubblmp}{\tens{\bar{u}}{_{\mathrm{BL}}^{\mu'}}}
\newcommand{\ubblnp}{\tens{\bar{u}}{_{\mathrm{BL}}^{\nu'}}}
\newcommand{\ubbla}{\tens{\bar{u}}{_{\mathrm{BL}}^0}}
\newcommand{\ubblb}{\tens{\bar{u}}{_{\mathrm{BL}}^1}}
\newcommand{\ubblc}{\tens{\bar{u}}{_{\mathrm{BL}}^2}}
\newcommand{\ubbld}{\tens{\bar{u}}{_{\mathrm{BL}}^3}}
\newcommand{\ubblm}{\tens{\bar{u}}{_{\mathrm{BL}}^\mu}}
\newcommand{\ubblal}{\tens{\bar{u}}{^{\mathrm{BL}}_0}}
\newcommand{\ubblbl}{\tens{\bar{u}}{^{\mathrm{BL}}_1}}
\newcommand{\ubblcl}{\tens{\bar{u}}{^{\mathrm{BL}}_2}}
\newcommand{\ubbldl}{\tens{\bar{u}}{^{\mathrm{BL}}_3}}

\newcommand{\athena}{\textsc{Athena\texttt{++}}}
\newcommand{\cosmos}{\textsc{Cosmos}}
\newcommand{\cosmospp}{\textsc{Cosmos\texttt{++}}}
\newcommand{\phantomcode}{\textsc{Phantom}}
\newcommand{\zeus}{\textsc{Zeus}}

\shorttitle{Tilted Disk Parameter Survey}
\shortauthors{C.~J.~White, E.~Quataert, O.~Blaes}

\begin{document}

% Title and author information
\title{Tilted Disks around Black Holes:\ A Numerical Parameter \\ Survey for Spin and Inclination Angle}
\author{Christopher~J.~White\altaffilmark{1}, Eliot~Quataert\altaffilmark{2}, and Omer~Blaes\altaffilmark{3}}
\altaffiltext{1}{Kavli Institute for Theoretical Physics, University of California Santa Barbara, Kohn Hall, Santa Barbara, CA 93107, USA}
\altaffiltext{2}{Department of Astronomy, University of California Berkeley, Campbell Hall, Berkeley, CA 94720, USA}
\altaffiltext{3}{Department of Physics, University of California Santa Barbara, Broida Hall, Santa Barbara, CA 93106, USA}

% Abstract
\begin{abstract}
  We conduct a systematic study of the properties of tilted accretion flows around spinning black holes, covering a range of tilt angles and black hole spins, using the general-relativistic magnetohydrodynamics code \athena. The same initial magnetized torus is evolved around black holes with spins ranging from $0$ to $0.9$, with inclinations ranging from $0^\circ$ to $24^\circ$. The tilted disks quickly reach a warped and twisted shape that rigidly precesses about the black hole spin axis with deformations in shape large enough to hinder the application of linear bending wave theory. Magnetized polar outflows form, oriented along the disk rotation axes. At sufficiently high inclinations a pair of standing shocks develops in the disks. These shocks dramatically affect the flow at small radii, driving angular momentum transport. At high spins they redirect material more effectively than they heat it, reducing the dissipation rate relative to the mass accretion rate and lowering the heating efficiency of the flow.
\end{abstract}

% Keywords
\keywords{accretion disks, black hole physics, hydrodynamics, MHD, relativity}

% Introduction
\section{Introduction}
\label{sec:introduction}

The general-relativistic (GR) effects of nonzero spin can have important consequences for accretion flows onto Kerr black holes. These effects are most dramatic when the spin of the black hole is not aligned with the rotational axis of the accretion disk, breaking the axisymmetry of the system. Such tilted disks are expected to occur in nature, for instance in supermassive black holes where the material falling from large radii in general has no reason to be aligned with the black hole spin. Lense--Thirring precession of a tilted flow is also a popular model for type~C low-frequency quasi-periodic oscillations in black hole X-ray binaries \citep{Ingram2009}.

First simulated in GR by \citet{Fragile2005} and \citet{Fragile2007} with the \cosmos{} and \cosmospp{} codes, tilted accretion disks display dynamical behavior not seen in untilted systems. For example, the monotonic relation between spin and the inner edge of the disk breaks down \citep{Fragile2009}, complicating black hole spin measurements. Even more noteworthy is the shape of tilted disks, as radial variation in the strength of Lense--Thirring precession results in disks warping and twisting as they precess around the black hole spin axis.

Simulations have shown that tilted disks can develop a pair of standing shocks that can alter the transport of angular momentum and dissipation of energy in the disk \citep{Fragile2008,Generozov2014}. Such structures may directly affect the variability observed in accreting systems \citep{Henisey2012}. Tilt also comes into play when relativistic jets are considered. For example, \citet{Polko2017} have studied the effect of relativistic jets on orienting tilted disks, and \citet{Liska2018} have examined how jets in such systems are oriented and how they precess.

The evolution of the shape of a tilted disk is often categorized as either as a diffusive process \citep{Bardeen1975,Papaloizou1983} or in terms of bending waves \citep{Papaloizou1995}, depending on whether the scale height $h/r$ is small or large compared to the effective $\alpha$-viscosity parameter. In the diffusive regime, the inner disk may align with the black hole spin, with the outer disk remaining misaligned. This will generally not happen in the bending wave regime, which is where most numerical work is done. Even when simulations can resolve geometrically thin accretion flows ($h/r \lesssim 0.1$), it can be prohibitively expensive to resolve flows so thin as to have $h/r < \alpha$.

\Citet{Krolik2015} contest that the distinction between diffusion and bending waves is based on isotropic viscosity models, and that in magnetohydrodynamics (MHD) the only important distinction is the degree of nonlinearity in the bending waves. As shown by \citet{Sorathia2013} and \citet{Krolik2014} using finite-volume \zeus{} simulations, MHD effects cannot be neglected in studying these disks. Even though magnetic forces on tilted disks are generally small compared to hydrodynamic ones, the anisotropy of MHD stresses and the turbulent nature of flows subject to the magnetorotational instability \citep[MRI,][]{Balbus1991} have important consequences for the transport of angular momentum, including the component of angular momentum responsible for alignment.

Even where numerical work has been done in the traditional bending-wave regime, there is a lack of consensus on what the final shape of the disk is. In contrast to the aforementioned authors, \citet{Nelson2000}, and later \citet{Nealon2015} and \citet{Nealon2016} using the smoothed-particle hydrodynamics \phantomcode{} code, find that disks can break, with fragments at different radii inclined relative to one another. This may be a result of the disk being unable to radially transport differentially applied Lense--Thirring torques rapidly enough. While direct comparison between such different numerical methods using the same physical conditions is difficult, the existence of these discrepancies between simulations indicates that the codes may not be in agreement regarding the rate of angular momentum transport relative to GR precession.

Neither \zeus{} nor \phantomcode{} employ GR directly, but rather add post-Newtonian corrections to capture the leading-order nodal (Lense--Thirring) and possibly apsidal (Einstein) precession terms, the magnitudes of which are sometimes artificially increased in order to observe the effects of differential precession in fewer orbital periods. \phantomcode{} is limited to simulating hydrodynamics with an artificial, isotropic viscosity, while the \zeus{} results employ MHD to self-consistently induce turbulence and drive angular momentum transport. As it is the interplay between GR dynamics and angular momentum transport that shapes the disk, these concerns need to be resolved. This motivates the use of a code that employs both GR and MHD, naturally including the post-Newtonian terms at all orders and directly modeling the MRI turbulence we know to be essential in accretion processes.

Other studies on tilted disk dynamics have gone so far as to employ full numerical relativity and thus capture the gravitational effect of the disk itself \citep{Mewes2016a,Mewes2016b}, though these are still done in viscous hydrodynamics rather than MHD. While the \cosmospp{} code uses both MHD and stationary GR, the aforementioned results were only able to marginally resolve the MRI. Moreover, those studies, like most of those done to date, examine one or two tilted configurations and compare them to an untilted control disk. As the shape and state of disks are apparently sensitive enough to GR effects and angular momentum transport that oversimplified approximations can lead to notably different outcomes, it is natural to ask how outcomes vary with \emph{physical} parameters, assuming numerics are handled adequately.

Here we more thoroughly explore a slice of parameter space, varying black hole spin and disk tilt angle, focusing on how these parameters affect the dynamics of tilted accretion disks. We use the finite-volume GR ideal MHD capabilities of \athena{} \citep{White2016} to evolve ten similar initial conditions with different combinations of spin and initial tilt. Our goal is to \emph{systematically} vary these parameters keeping all else fixed, so that we may quantify how the accretion flows depend on these important variables. At the same time, we want to be sure that we are including all relevant physics (GR and MHD) in order to accurately simulate the behavior of tilted disks.

The simulations are described in \S\ref{sec:setup}, with important definitions regarding tilted coordinate systems given in \S\ref{sec:measuring}. The results are analyzed in terms of the shape of the disk midplane (\S\ref{sec:shape}), the resulting poloidal structure (\S\ref{sec:poloidal}), and the standing shocks we find (\S\ref{sec:shocks}). Concluding remarks are made in \S\ref{sec:discussion}. As these are non-radiative, non-self-gravitating, ideal MHD simulations, the results scale to any black hole mass $M$, with all lengths implicitly in units of $GM/c^2$ and times in units of $GM/c^3$.

% Numerical setup
\section{Numerical Setup}
\label{sec:setup}

We choose to use spherical Kerr--Schild coordinates \citep{Font1998} aligned with the black hole axis. Our coordinates cover the entire sphere, including the appropriate transmissive polar boundary, and extend from inside the horizon to $r = 70$. In order to keep the timesteps reasonable, we use static mesh refinement to keep the part of the grid near the poles at low resolution. The regions within $22.5^\circ$ of the poles are at an effective resolution of $56 \times 32 \times 64$ in $r$, $\theta$, $\phi$; regions between $22.5^\circ$ and $33.75^\circ$ from the poles are at $112 \times 64 \times 128$; and the grid within $56.25^\circ$ of the midplane is at $224 \times 128 \times 256$. For comparison, the fiducial resolution of \citet{Fragile2007} was $128^3$ (with the outer radius of the simulation at $r = 120$). Improvement of the azimuthal resolution in particular means that at late times we achieve a relativistic MRI quality factor
\begin{equation}
  Q_\phi \equiv \frac{2\pi (r^{3/2} + a)}{\sqrt{g_{33}} \Delta\phi} \paren[\bigg]{\frac{b_3 b^3}{\rho + \Gamma/(\Gamma-1) \times \pgas + 2 \pmag}}^{1/2}
\end{equation}
in the inner parts of the disk ranging from approximately $70$ (no spin) to $90$ (highest spin). \Citet{Hawley2011} suggest convergence sets in around values of $15$ to $20$. We note that the results of \S\ref{sec:shocks} indicate that while the MRI may be necessary to have any accretion, once the flow develops standing shocks it is the adequate resolution of these \emph{hydrodynamic} features that may become more important.

In choosing to incline the disk relative to the grid, rather than have the two be initially aligned, we are limited in how extreme a tilt we can achieve before the disk lies in the lower resolution polar region. However, this choice means the disk is free to precess about the spin axis without approaching the grid axis. For example, if we had the disk and grid be initially aligned, and the spin axis misaligned from these by $24^\circ$, then after half a precession the disk would still be $24^\circ$ off the spin axis but $180^\circ$ out of phase with the grid, with the result being the disk and grid misaligned by $48^\circ$ and much of the disk worryingly close to the grid polar axis. Even introducing a global precession into the grid itself will not solve this, given that the disk can precess differently at different radii.

\athena{} is run with a second-order van~Leer integrator at a Courant--Friedrichs--Lewy number of $0.3$, where $1/3$ is the maximum for this integrator in $3$ spatial dimensions. We use the second-order modified van~Leer spatial reconstruction from \citet{Mignone2014}. The HLLE Riemann solver is used to calculate fluxes.

The initial conditions are those of the hydrostatic equilibrium solution of \cite{Fishbone1976} with inner edge at $r = 15$, pressure maximum at $r = 25$, adiabatic index of $\Gamma = 4/3$, and peak rest mass density of $\rho = 1$. These are the same inner edge and pressure maximum values as in \citet{Fragile2007}, though we have a more compressible adiabatic index compared to their $\Gamma = 5/3$. These are thick tori with no cooling added, so at late times the scale height
\begin{equation}
  h/r \equiv \frac{\int \abs{\theta' - \pi/2} \rho \sqrt{-g} \, \dth\,\dph}{\int \rho \sqrt{-g} \, \dth\,\dph},
\end{equation}
ranges from $0.1$ to $0.2$ inside $r = 20$. This solution is calculated for an untilted disk and is then given the desired inclination (so it is no longer exactly in equilibrium).

We next add a magnetic field from a vector potential that has only an azimuthal component in disk-aligned coordinates proportional to $\max(\rho - 0.2, 0)$. The strength of the magnetic field is normalized with respect to plasma $\beta^{-1} \equiv \pmag / \pgas$ such that
\begin{equation}
  \ave{\beta^{-1}} \equiv \frac{\int \beta^{-1} \rho \sqrt{-g} \, \dr\,\dth\,\dph}{\int \rho \sqrt{-g} \, \dr\,\dth\,\dph}
\end{equation}
is $10^{-3}$.

We perform ten simulations out to a time of $t = 4000$, with snapshots made every $10$ time units. Two of these are around a Schwarzschild black hole, one with an initial disk inclination of $0^\circ$, and another with an inclination of $24^\circ$ used to verify that the results do not depend on the orientation of the disk with respect to the grid. We also run simulations with $0^\circ$, $8^\circ$, $16^\circ$, and $24^\circ$ inclinations around black holes with dimensionless spins $a = 0.5$ and $a = 0.9$. Steady-state inflow is established out to approximately $r = 10$ by the end of the simulations.

% Measuring disk properties
\section{Measuring Disk Properties}
\label{sec:measuring}

In order to analyze the properties of a tilted accretion disk, we must have a definition of the disk's orientation at a moment in time, either globally or as a function of radial coordinate. We choose the gas angular momentum defined to be
\begin{equation} \label{eq:l}
  \tens{L}{^{\hat{\imath}}} \equiv [i\ j\ k] \tens{r}{^{\hat{\jmath}}} \tensor*{T}{_{\mathrm{gas}}^{0\hat{k}}},
\end{equation}
where hats indicate Cartesian coordinates related to our underlying spherical coordinates in the usual way, the brackets denote the antisymmetric Levi--Civita symbol, and
\begin{equation}
  \tens{T}{_{\mathrm{gas}}^{\mu\nu}} \equiv \paren[\bigg]{\rho + \frac{\Gamma}{\Gamma-1} \pgas} \tens{u}{^\mu} \tens{u}{^\nu} + \pgas \tens{g}{^{\mu\nu}}
\end{equation}
are the components of the hydrodynamic stress-energy tensor. The results are much the same if instead of $T_\mathrm{gas}^{0\mu}$ we choose $\rho u^0 u^\mu$ or even $\rho u^\mu / u^0$ as is done by some authors.

From the angular momentum we can construct the Euler angles $i$ (inclination, sometimes called $\beta$ in the literature) and $\phi_0$ ($90^\circ$ less than the longitude of ascending node, sometimes called $\alpha$). They are given by
\begin{subequations} \label{eq:euler} \begin{align}
  i & = \cos^{-1}\paren[\bigg]{\frac{L^z}{((L^x)^2 + (L^y)^2 + (L^z)^2)^{1/2}}}, \label{eq:euler:i} \\
  \phi_0 & = \tan^{-1}(L^y, L^x). \label{eq:euler:phi0}
\end{align} \end{subequations}
These angles relate the coordinates $\theta,\phi$ aligned with the grid and black hole to the coordinates $\theta',\phi'$ aligned with the disk according to
\begin{subequations} \label{eq:spherical_transformation} \begin{align}
  \theta' & = \cos^{-1}(\cos i \cos\theta + \sin i \cos\phi_0 \sin\theta \cos\phi + \sin i \sin\phi_0 \sin\theta \sin\phi), \\
  \phi' & = \tan^{-1}(-\sin\phi_0 \sin\theta \cos\phi + \cos\phi_0 \sin\theta \sin\phi, \notag \\
  & \quad \qquad -\sin i \cos\theta + \cos i \cos\phi_0 \sin\theta \cos\phi + \cos i \sin\phi_0 \sin\theta \sin\phi), \\
  \theta & = \cos^{-1}(-\sin i \sin\theta' \cos\phi' + \cos i \cos\theta'), \\
  \phi & = \tan^{-1}(\sin i \sin\phi_0 \cos\theta' + \cos i \sin\phi_0 \sin\theta' \cos\phi' + \cos\phi_0 \sin\theta' \sin\phi', \notag \\
  & \quad \qquad \sin i \cos\phi_0 \cos\theta' + \cos i \cos\phi_0 \sin\theta' \cos\phi' - \sin\phi_0 \sin\theta' \sin\phi').
\end{align} \end{subequations}
Note that for a circular geodesic, $i$ is not exactly the maximum deviation of $\theta$ from the midplane over the orbit. However, these two senses of inclination are in close agreement, never deviating by more than approximately a degree over all radii in the range of spins and tilts considered.

% Disk shape
\section{Disk Shape}
\label{sec:shape}

In each of our simulations, the disk orientation as a function of radius stops evolving by $t = 1500$, except for a slow, rigid-body precession. This precession rate is $2.4^\circ$ per $1000$ time units in the $a = 0.5$ cases, and $3.8\text{--}4.1^\circ$ per $1000$ time units in the $a = 0.9$ cases. For comparison, \citet{Fragile2007}, whose initial conditions match ours in terms of the inner edge and pressure maximum radii, find a rigid-body precession period of approximately $0.3\ (M/\msun)\ \mathrm{s}$ for $a = 0.9$ with a $15^\circ$ inclination. This corresponds to $6^\circ$ per $1000$ time units. As with \citeauthor{Fragile2007}, our precession rate is consistent with the total angular momentum of the rigid body and the Lense--Thirring torque applied to it. As a measure of conservation of angular momentum when the disk and grid are not aligned, the $24^\circ$ inclined disk around the Schwarzschild black hole precesses less than $0.1^\circ$ per $1000$ time units.

\begin{figure*}
  \centering
  \includegraphics{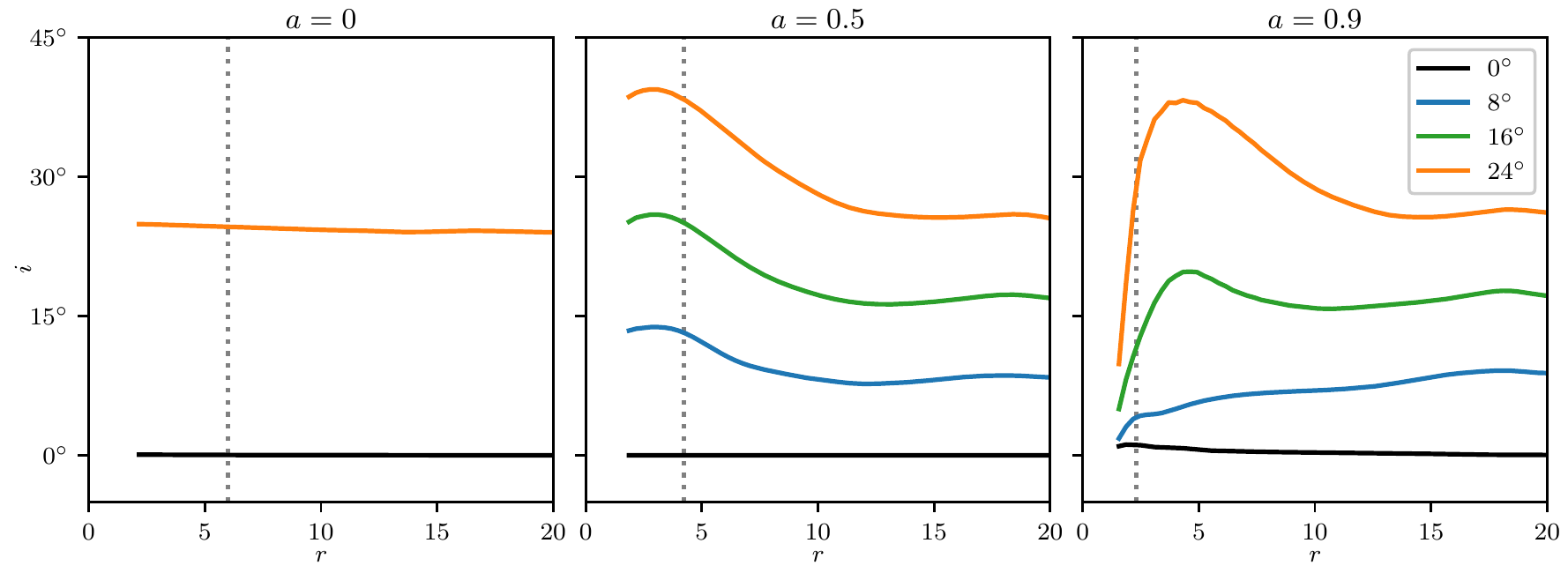}
  \caption{Inclinations of disks as functions of radius, time-averaged over $3000 \leq t \leq 4000$. The flatness of the $a = 0$ and $0^\circ$ lines is expected, indicating conservation of angular momentum. Of the six spinning, tilted cases, five display a peak in inclination at small radii. This shows the disks warping further away from alignment with the black hole at small radii compared to large radii, before beginning to align at very small radii. The vertical dotted lines indicate the radius of the equatorial ISCO. \label{fig:inclination_measured}}
\end{figure*}

\begin{figure*}
  \centering
  \includegraphics{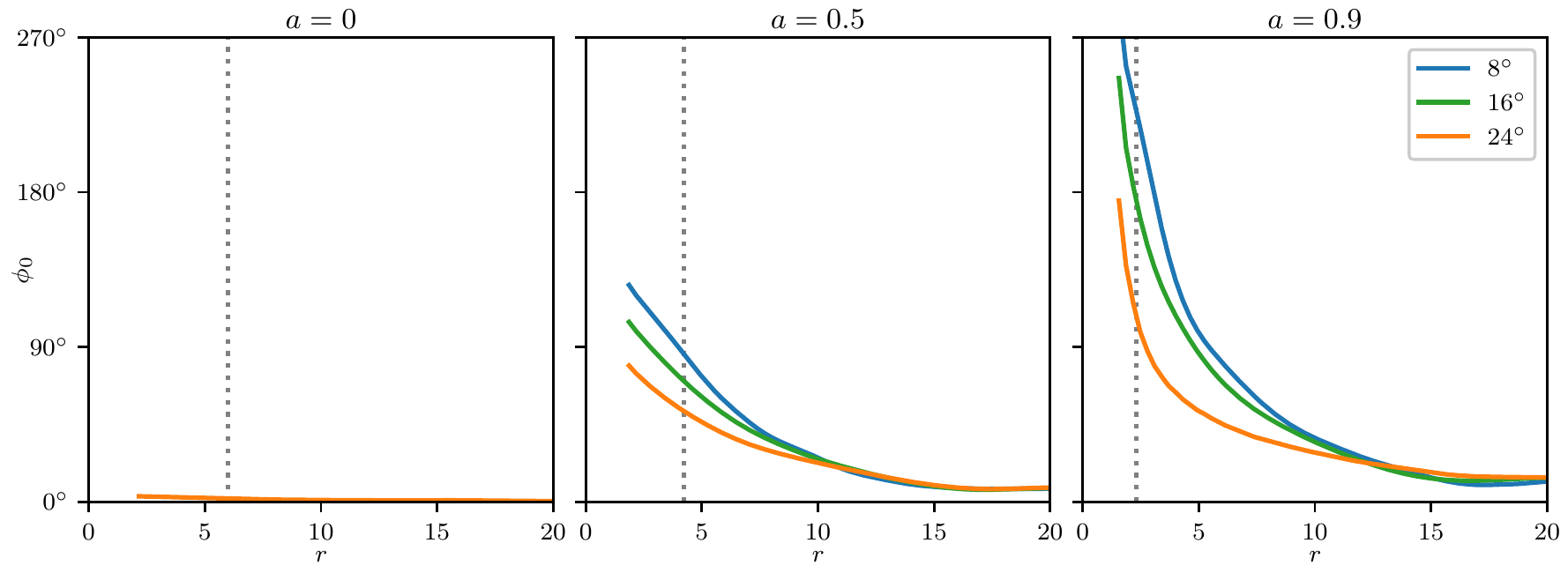}
  \caption{Precession angles of disks as functions of radius, time-averaged over $3000 \leq t \leq 4000$. The flatness of the line for the tilted Schwarzschild case (left panel) reflects conservation of angular momentum. The other six cases show the disk twisting a significant fraction of a full circle as material moves inward. The vertical dotted lines indicate the radius of the equatorial ISCO. The untilted disk runs are omitted, as $\phi_0$ is not well defined for $i = 0$. \label{fig:precession_measured}}
\end{figure*}

The radially dependent orientation defined by the Euler angles $i$ and $\phi_0$, averaged over $3000 \leq t \leq 4000$, is shown in Figures~\ref{fig:inclination_measured} and~\ref{fig:precession_measured}. We mark the equatorial innermost stable circular orbit (ISCO) for reference. As the disk is rotating rigidly at a rate much slower than the precession of geodesics at small radii, we can time-average $\phi_0$ without worrying that we are averaging over many full precession periods at these radii. In fact, over the period of the time averaging, the solid body rotates less than $10^\circ$ at all radii.

The small deviation in $i$ of the initially untilted disk with $a = 0.9$ (Figure~\ref{fig:inclination_measured}, right panel, black line) measures the imprecision of this definition due to the stochastic nature of the accretion flow, as well as magnetic field angular momentum not included in \eqref{eq:l}. The relative flatness of the initial $24^\circ$ inclination when there is no spin (left panel, colored line) demonstrates that the code does well at conserving angular momentum even when the flow is not aligned with the grid or coordinate system.

The most striking feature of Figure~\ref{fig:inclination_measured} is that the inclinations increase with decreasing radius. That is, the disks are warped even further out of the plane near the black hole. Only at the highest spin and lowest inclination is this not seen. This additional warping is observed in other simulations of tilted thick disks; see for example Figure~12 of \citet{Fragile2007} or Figure~7 of \citet{Mewes2016b}.

One way this effect can arise is if angular momentum exchange between adjacent radii occurs primarily at the ascending and descending nodes. (See \S\ref{sec:shocks} for a description of the standing shocks found in our simulation, whose locations near the nodes can cause this to happen.) The inclination angle defined in \eqref{eq:euler:i} is notably different from the inclination of the orbit's direction while crossing the line of nodes, $\pm\tan^{-1}(u^2 / u^3)$. In particular, for a constant inclination as defined by angular momentum (or by antinodal latitude), this nodal inclination decreases with decreasing $r$. Conversely, if nodal inclination is kept constant with radius, the angular momentum inclination of circular geodesics must increase toward the black hole. Thus if the stress responsible for transporting angular momentum tends to align the trajectories of adjacent fluid elements, and if this mechanism operates primarily near the line of nodes, one expects to see inclination increasing toward smaller radii in steady state.

With Figure~\ref{fig:precession_measured} we again have a measure of the numerical non-conservation of angular momentum, given by the small deviation of the $24^\circ$ inclined disk around the Schwarzschild black hole (left panel, colored line). For the inclined flows around spinning black holes, we see the disks being twisted by differential precession. This effect is strong enough in the $a = 0.9$ case that the very inner parts of the disk can be twisted beyond $180^\circ$ relative to the outer parts.

While we observe a large amount of differential precession in these simulations, we note that the effect would be much stronger if not for internal stresses redistributing angular momentum. To see this, we consider a model in which particles orbit on circular geodesics, accumulating a precession angle at a rate $\dd\phi_0/\dtau$ with respect to their proper time (see the appendix, \S\ref{sec:nodal}). This can be converted to a differential precession $\dd\phi_0/\dr$ by dividing by a prescribed radial velocity $u^1 \equiv \dr/\dtau$.

\begin{figure*}
  \centering
  \includegraphics{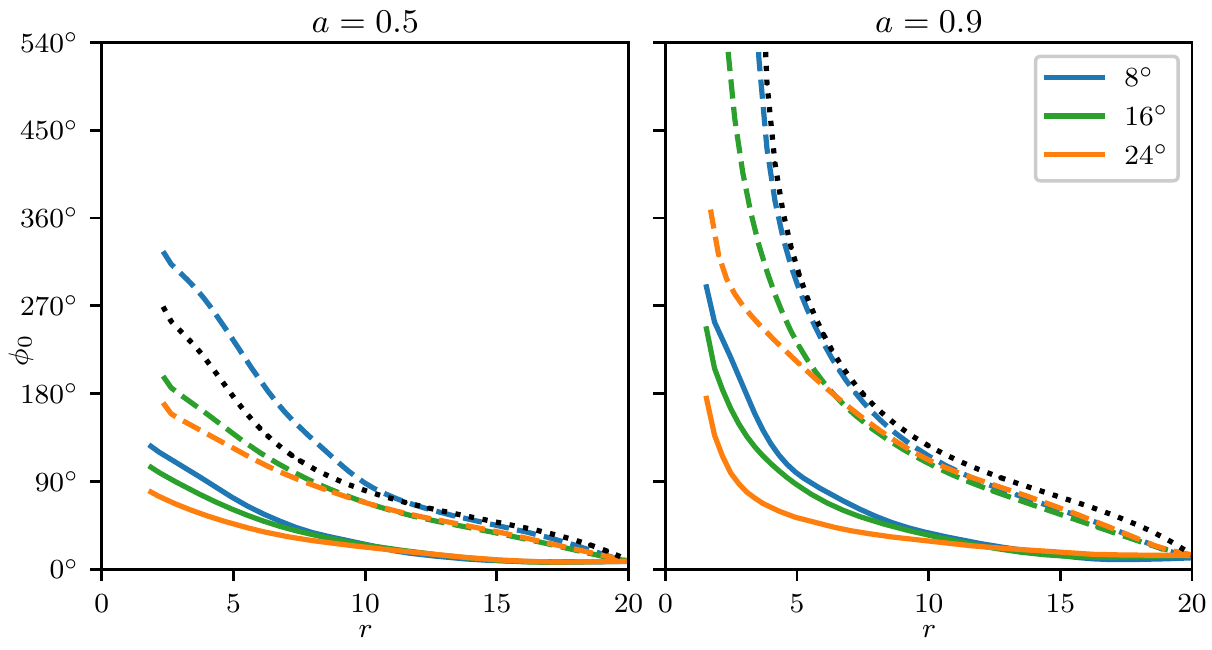}
  \caption{Precession angles of disks as functions of radius, time-averaged over $3000 \leq t \leq 4000$. Solid lines show the same data as in Figure~\ref{fig:precession_measured}. Dashed lines show the prediction given just nodal precession of geodesics and the simulations' measured radial velocities. The dotted line is the prediction if the corresponding untilted simulation's radial velocity is used instead. The displacement between solid and broken lines comes from redistribution of angular momentum from internal disk torques. The spread in dashed lines relative to the dotted lines shows the effect of different radial velocities in determining disk shape. That this spread roughly agrees with the spread in solid lines indicates that the differences in disk shape among disks with different inclinations can be traced to the different infall velocities in those simulations. \label{fig:precession_geodesic}}
\end{figure*}

We show the results of this calculation in Figure~\ref{fig:precession_geodesic}. The solid lines are the observed precession angles as shown in Figure~\ref{fig:precession_measured}. The dashed lines are computed by integrating $\dd\phi_0/\dr$ inward from $r = 20$, where the curves are matched to the observed values. The values of $u^1$ used are taken from the respective simulations; they are density-weighted shell averages within one scale height of the disk midplane and averaged in time over $3000 \leq t \leq 4000$. The fact that the dashed lines are higher than the solid lines indicate the disk is transporting angular momentum induced by the Lense--Thirring effect from small radii to large radii.

The spread in the observed $\phi_0$ profiles matches that of this geodesic calculation, and this is due almost entirely to differences in $u^1$ between simulations at different inclinations. That is, the differences in differential precession between simulations with the same value of $a$ are dominated by differences in radial velocity, not by minute differences in precession of geodesics at different inclinations. While nodal precession of circular geodesics does have a small dependence on inclination (beyond the linear order used by Lense and Thirring), this effect is hardly noticeable at the scale of the plot. Instead of using the radial velocities from each simulation we can use the values from the untilted simulation at the appropriate spin, thus removing the effect of different $u^1$ values, and the result is shown as the black dotted line in Figure~\ref{fig:precession_geodesic}.

As there are internal aligning and precessing torques in these thick disks, one might consider comparing our results to the theory of bending waves. Convenient formulas for the solution to the linearized bending wave problem are given in \citet{Foucart2014}. This framework is developed in a Newtonian context with a nonspherical potential. The solution depends on the orbital, epicyclic, and vertical frequencies of orbits, as well as the integrated density and pressure of the disk and an effective viscosity for the disk.

\begin{figure*}
  \centering
  \includegraphics{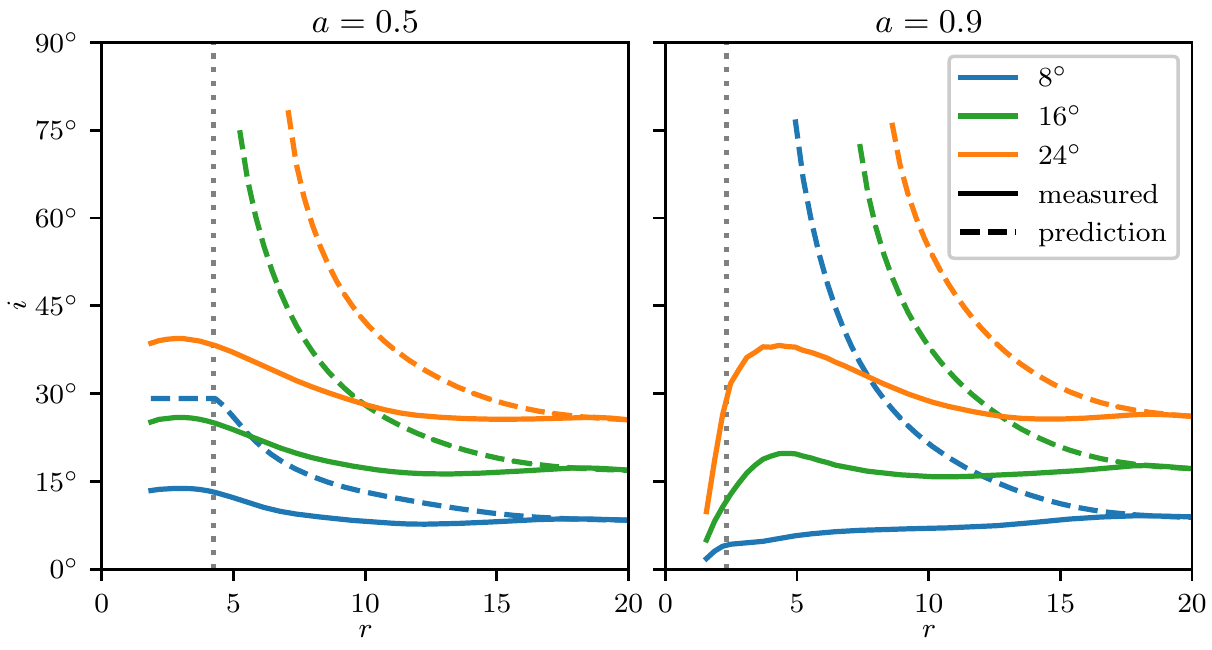}
  \caption{Directly measured inclinations (solid, time-averaged over $3000 \leq t \leq 4000$), and those predicted by linear bending wave theory (dashed, computed from time-averaged values of surface density and pressure). The linear solutions are matched onto the observed values at $r = 20$. The disagreement interior to this indicates our disks are not well described by the linear theory we employ. The vertical dotted lines indicate the radius of the equatorial ISCO. \label{fig:inclination_linear}}
\end{figure*}

The result of this comparison, however, only highlights the inapplicability of linear bending wave theory to most realistic tilted disks around black holes. The theory makes a number of assumptions that are violated here:\ that the epicyclic and vertical frequencies deviate only slightly from their Keplerian values (see the appendix, \S\ref{sec:frequencies}, for the definitions and plots of these in Kerr spacetime), that there is no stress at the ISCO, and that $h/r \gg \alpha$. In order to illustrate this incompatibility, we show the observed and theoretical disk inclinations in Figure~\ref{fig:inclination_linear}.

\begin{figure*}
  \centering
  \includegraphics{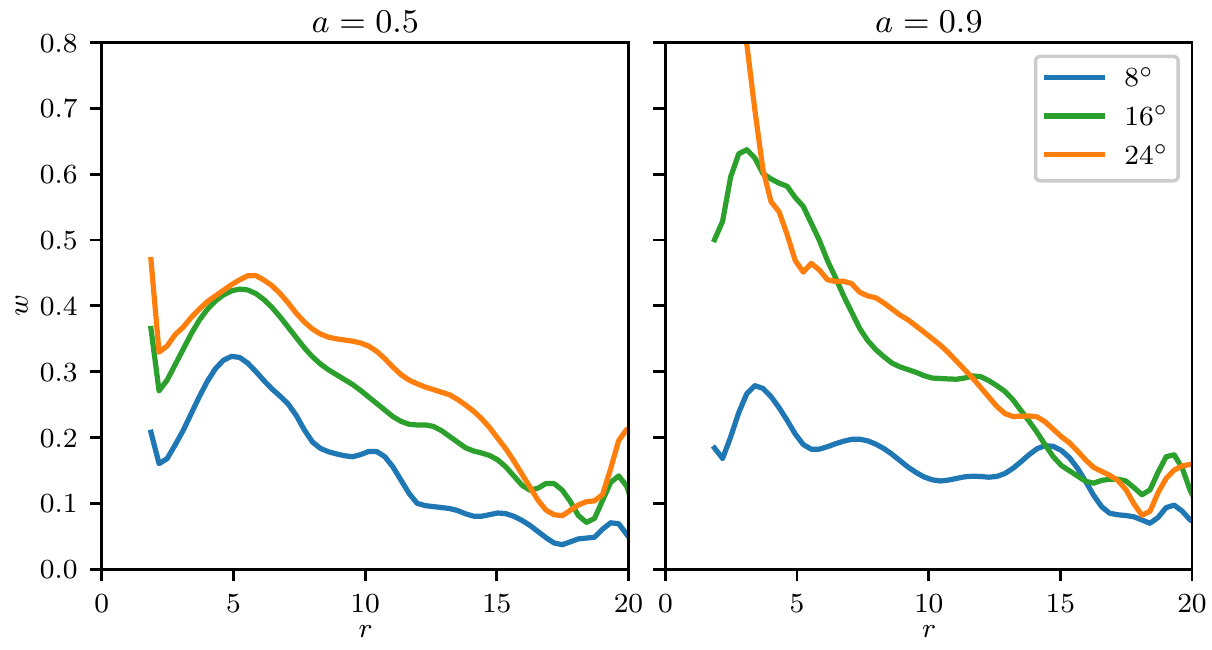}
  \caption{Dimensionless warps as functions of radius, time-averaged over $3000 \leq t \leq 4000$. Warping increases at small radii, and the effect is stronger in the simulations with higher inclinations. \label{fig:shape_warp}}
\end{figure*}

\begin{figure*}
  \centering
  \includegraphics{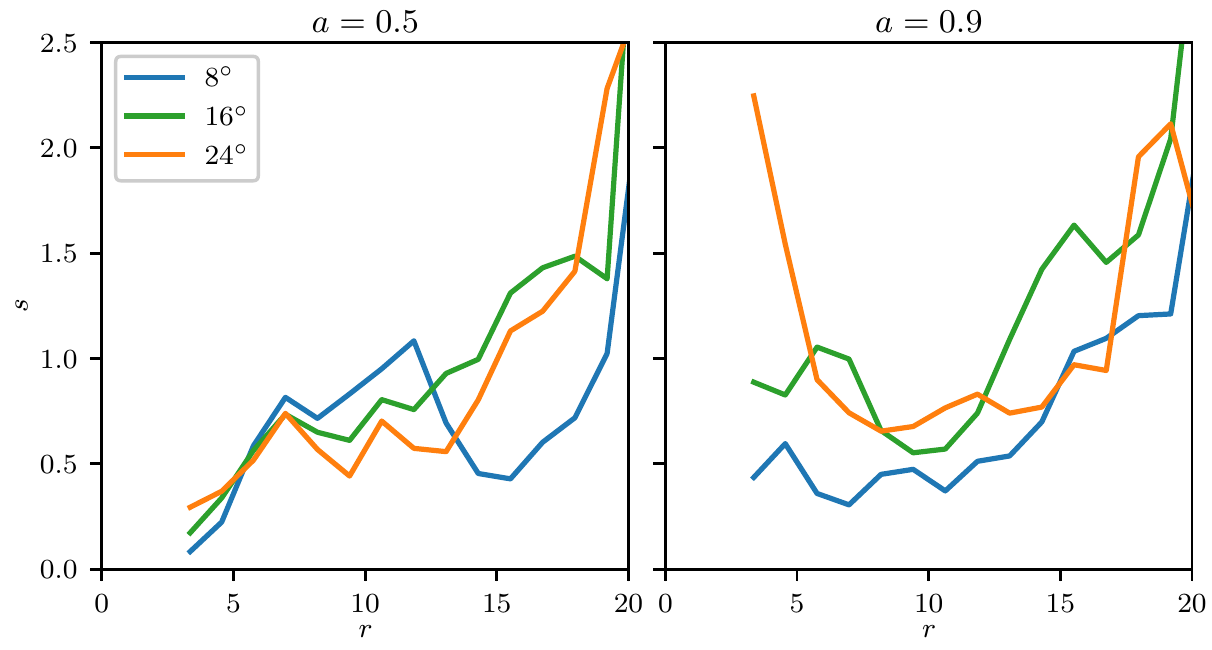}
  \caption{Dimensionless twists as functions of radius, time-averaged over $3000 \leq t \leq 4000$. Twist generally decreases with decreasing radius, and it does not show a strong dependence on either spin or inclination. \label{fig:shape_twist}}
\end{figure*}

We can quantify the out-of-plane bending of the disk, defining the dimensionless warp and twist to be
\begin{align}
  w & \equiv r \abs[\bigg]{\vec{n} \times \frac{\dd\vec{n}}{\dr}}, \\
  s & \equiv r^2 \abs[\bigg]{\frac{\dd^2\vec{n}}{\dr^2} \times \paren[\bigg]{\vec{n} \times \frac{\dd\vec{n}}{\dr}}} \cdot \abs[\bigg]{\frac{\dd\vec{n}}{\dr}}^{-1}.
\end{align}
Here $\vec{n}$ is the Cartesian vector of angular momentum components $L^{\hat{\imath}}$, scaled to have unit norm. These are similar to the definitions given in \citet{Shiokawa2013}, but we use the radial coordinate rather than scale height to make them dimensionless. The nontrivial warps and twists from our simulations are plotted in Figures~\ref{fig:shape_warp} and~\ref{fig:shape_twist}.

% Poloidal structure
\section{Poloidal Structure}
\label{sec:poloidal}

Here we examine the structure of our tilted accretion flows in the poloidal plane. For the plots in this section we slice the simulation along the plane orthogonal to the line of nodes, orienting all images such that the black hole spin axis is vertical. Thus the disks will always be tilted down to the right. These slices are averaged in time for $3000 \leq t \leq 4000$. Streamlines are constructed from the $r$-~and $\theta$-components of vector quantities. In all cases we show the inner $40 \times 40$ gravitational radius patch of the simulation, with the black hole masked in the center.

\begin{figure*}
  \centering
  \includegraphics{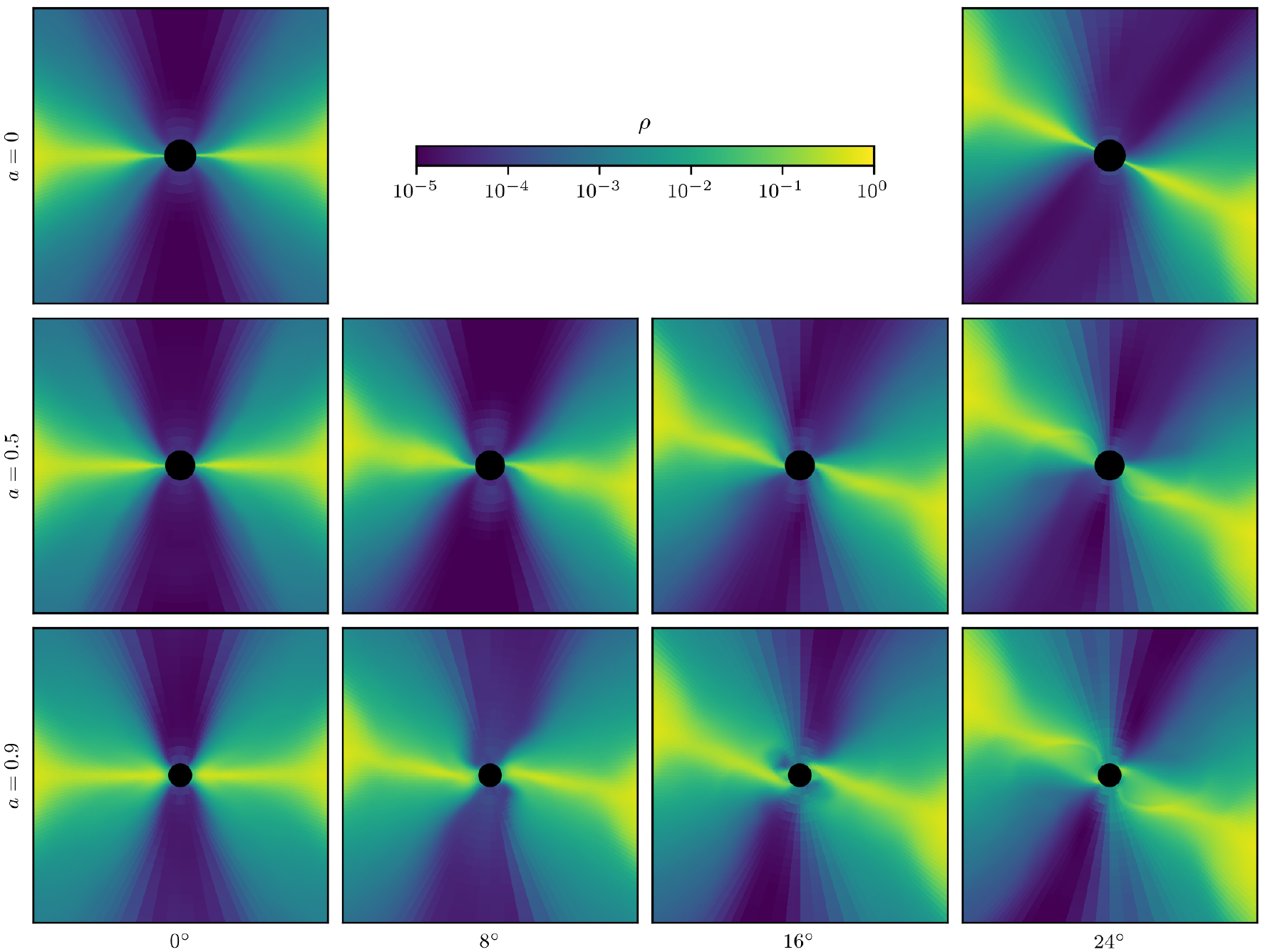}
  \caption{Time-averaged poloidal slices of density for all the simulations. Each panel is $40$ gravitational radii wide, with the black hole spin pointing up and the line of ascending nodes receding into the page. The two upper panels should look like tilted versions of one another, since there is no spin to break spherical symmetry. At the highest spins and inclinations the warping at small radii (see Figure~\ref{fig:inclination_measured}) can be seen directly, with the disk taking on a complex shape. \label{fig:vertical_rho}}
\end{figure*}

Slices of rest-frame density $\rho$ are shown in Figure~\ref{fig:vertical_rho}. From these it is clear that most of the disks display large bending out of their original planes. For the highest inclinations, the disk even appears to split into two streams:\ a thinner one close to vertical in the image, and a broader one. The fact that these streams remain clearly visible even after the long time averaging indicates they are not transient phenomena.

\begin{figure*}
  \centering
  \includegraphics{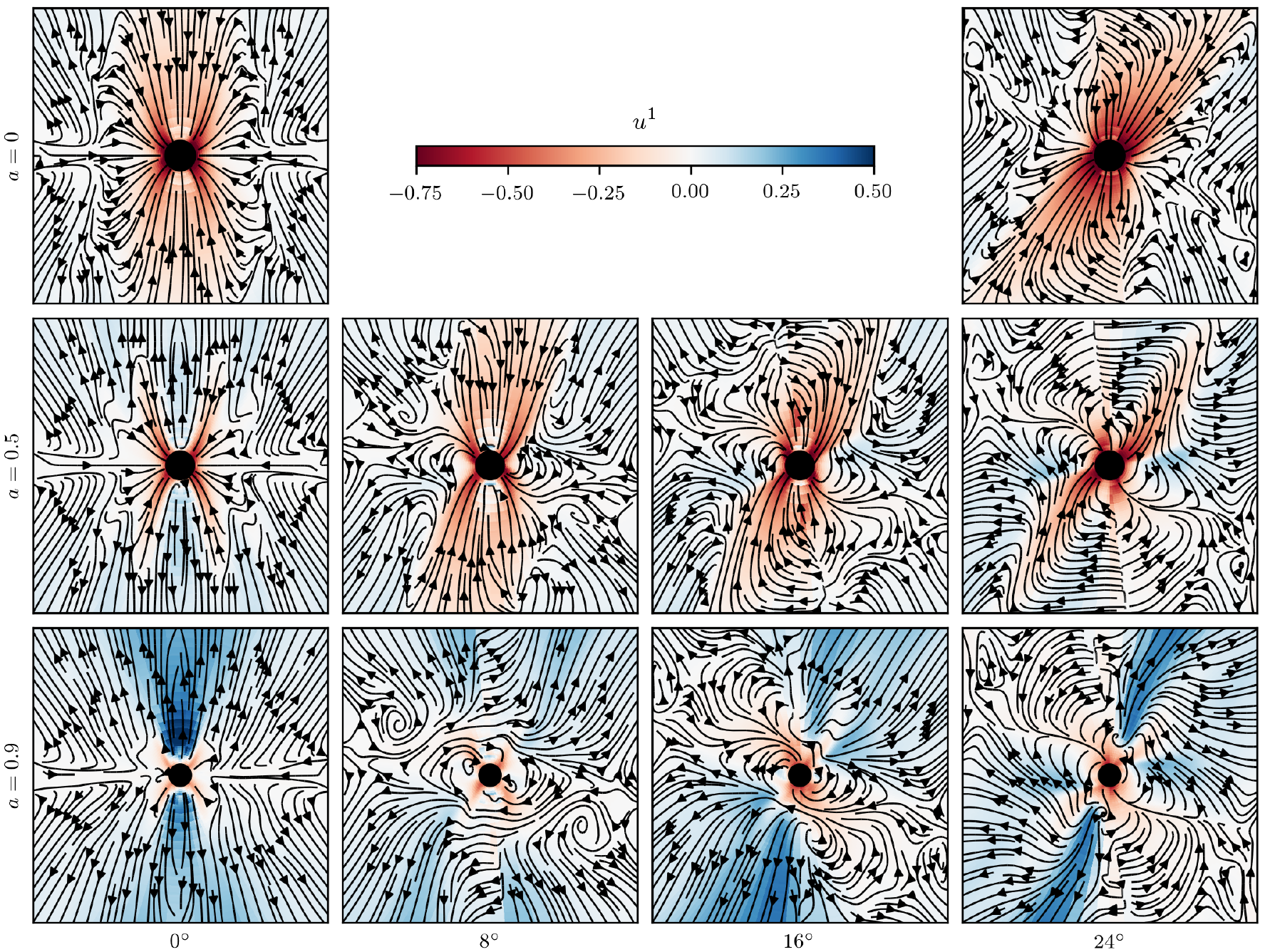}
  \caption{Time-averaged poloidal slices of radial velocity for all the simulations. The streamlines indicate the velocity field in the plane of the figure. Each panel is $40$ gravitational radii wide, with the black hole spin pointing up and the line of ascending nodes receding into the page, as in Figure~\ref{fig:vertical_rho}. The $a = 0.9$ simulations tend to develop a strong outflow that is oriented with the disk, though low but nonzero disk inclination seems to suppress the outflow overall. A pattern of positive radial velocity above the disk and negative below, reversed on the opposite side of the image, can be seen at high inclinations. This pattern matches that of Figure~1 of \citet{Fragile2008}, where it is explained via ordered eccentric orbits. \label{fig:vertical_u1}}
\end{figure*}

The density plots indicate that an evacuated, low-density region forms along the direction of the disk's angular momentum rather than parallel to the black hole spin. Untilted simulations generically see outflows along the common axis, so this raises the question of what direction the material is flowing in the tilted case. For this reason we construct Figure~\ref{fig:vertical_u1} showing the velocity streamlines of $u^1$ and $u^2$, with colors indicating the radial velocity.

In the untilted cases, we clearly see the polar outflow strengthening with increasing spin. Comparing tilted disks to untilted disks at a given nonzero spin, we see that this outflow is overall suppressed in the former. Still, its direction is generally aligned with the low-density regions.

By comparing the lower right panels in Figures~\ref{fig:vertical_rho} and~\ref{fig:vertical_u1}, we can analyze the velocity structure in relation to the disk location. In particular, there is a pattern of positive $u^1$ above the disk on the right-hand side and below the disk on the left, with negative $u^1$ in the opposite positions. The outward radial velocity is not an unbound outflow, but rather the part of an eccentric orbit going from periapsis to apoapsis. Above the disk, the matter is orbiting with apoapsis behind the figure and periapsis in front, with the reverse holding below the disk. This pattern persists through time averaging, showing that it is fixed in place relative to the rigidly precessing disk. These eccentric orbits, with those above the disk $180^\circ$ out of phase relative to those below and with apsides near to the disk's line of nodes, are observed in \citet[cf.\ Figure~1]{Fragile2008}, where they are found to be signatures of standing shocks. The coherence of the velocity pattern decreases with decreasing spin and tilt, so that we expect shocks to be strongly influencing the accretion flows only in our simulations with higher tilts and spins.

\begin{figure*}
  \centering
  \includegraphics{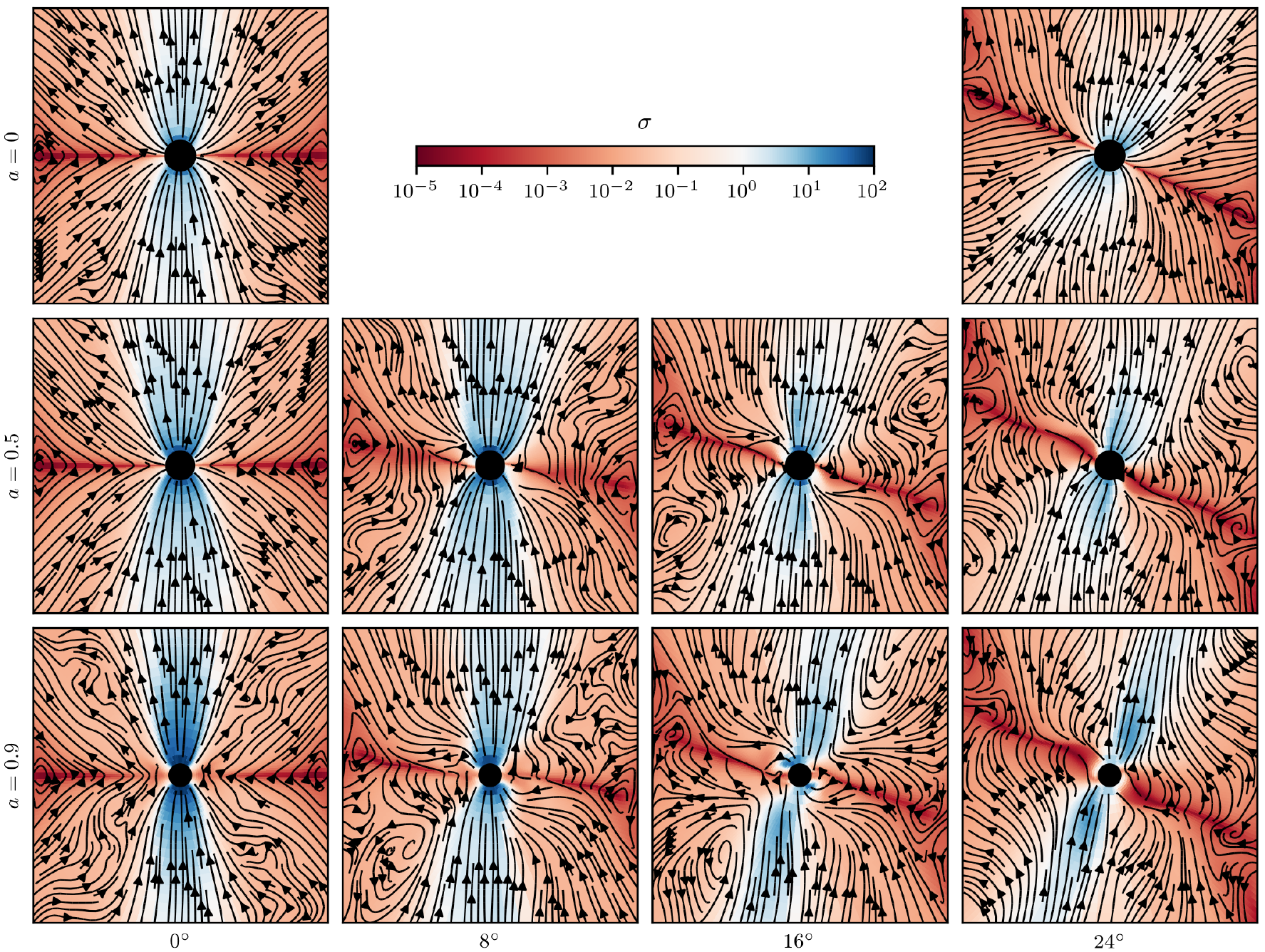}
  \caption{Time-averaged poloidal slices of magnetization for all the simulations. The streamlines indicate the magnetic field in the plane of the figure. Each panel is $40$ gravitational radii wide, with the black hole spin pointing up and the line of ascending nodes receding into the page, as in Figures~\ref{fig:vertical_rho} and~\ref{fig:vertical_u1}. In all cases there is a low magnetization current sheet that, on average, tracks the midplane of the disk, as well a high magnetization outflow oriented perpendicular to the disk. \label{fig:vertical_sigma}}
\end{figure*}

Also worth viewing in poloidal slices is the magnetization of the material. For this we choose $\sigma \equiv 2 \pmag / \rho$, shown in Figure~\ref{fig:vertical_sigma}. Here streamlines are constructed from $B^1$ and $B^2$. In a time-averaged sense, there is a current sheet in a relatively narrow region roughly aligned with the midplane of the disk. A radial current is expected whenever a disk with a poloidal magnetic field rotates faster at its midplane, shearing the field azimuthally with a reversal at the midplane. Tilting the disk does not change this general feature except by warping the current sheet.

Unsurprisingly, the low-density polar regions seen in Figure~\ref{fig:vertical_rho} correspond to the higher magnetizations in Figure~\ref{fig:vertical_sigma}. It is more noteworthy that the average magnetic field orientation in this region appears to have less inclination to the black hole spin axis (vertical in the figures) than the disk's rotational axis does. This is despite the initial field being aligned with the disk, and in contrast with the average velocity structure in the same regions, which is close to perpendicular to the plane of the disk. Thus the velocity and magnetic structures of a jet may carry information about the relative inclination of the disk to the black hole.

Note these are only weak polar outflows, as expected given that we did not use initial conditions designed to yield a magnetically arrested disk (MAD). Define
\begin{subequations} \begin{align}
  \dot{M} & \equiv -\oint \rho u^1 \sqrt{-g} \, \dth\,\dph, \\
  \Phi & \equiv \frac{1}{2} \oint \sqrt{4\pi} \abs{B^1} \sqrt{-g} \, \dth\,\dph, \\
  \varphi & \equiv \frac{\Phi}{M \ave{\dot{M}}^{1/2}},
\end{align} \end{subequations}
where the time average in the last equation is taken over $3000 \leq t \leq 4000$. We find $\varphi \lesssim 8$ in all our simulations, whereas the prototypical strong-jet-producing MAD simulations of \citet{Tchekhovskoy2011} have $\varphi \approx 47$.

% Standing shocks
\section{Standing Shocks}
\label{sec:shocks}

In order to investigate standing shocks, we first construct averages of density, gas pressure, and velocity over the time interval $3000 \leq t \leq 4000$, rotating each snapshot in $\phi$ in order to compensate for the global precession rate. The quantities other than density are density weighted in this average. At each point in this average state, assumed to be stationary, we compute the nonadiabatic source term for generation of entropy per unit coordinate volume per unit coordinate time, $\rho u^i \partial_i s$, where $s \equiv \log(\pgas / \rho^\Gamma) / (\Gamma-1)$. This follows from writing a ``conservation law'' for entropy per unit volume analogous to that for mass per unit volume:
\begin{equation}
  \nabla_\mu(s \rho u^\mu) = s \nabla_\mu(\rho u^\mu) + \rho u^\mu \nabla_\mu s = \rho u^\mu \partial_\mu s.
\end{equation}
The second equality follows from mass conservation $\nabla_\mu(\rho u^\mu) = 0$. The right-hand side must be the nonadiabatic entropy source per unit time and volume in the rest frame of the fluid, whose direct evaluation is normally hindered by not knowing $\partial_0 s$ given only $s$ at a single time slice. However in steady state this term must vanish, and we are left with the source term $\rho u^i \partial_i s$. Dividing by $u^0$ converts the rate from that measured in the fluid frame to that measured with respect to Kerr--Schild time $t$, but we must also multiply by $u^0$ to obtain a rate per unit coordinate volume rather than per unit fluid-frame volume. Multiplying by temperature $T \equiv \pgas / \rho$ gives the irreversible heating rate per unit coordinate time per unit coordinate volume $q \equiv \pgas u^i \partial_i s$. This term should have large positive values in regions of shocks.

\begin{figure*}
  \centering
  \includegraphics{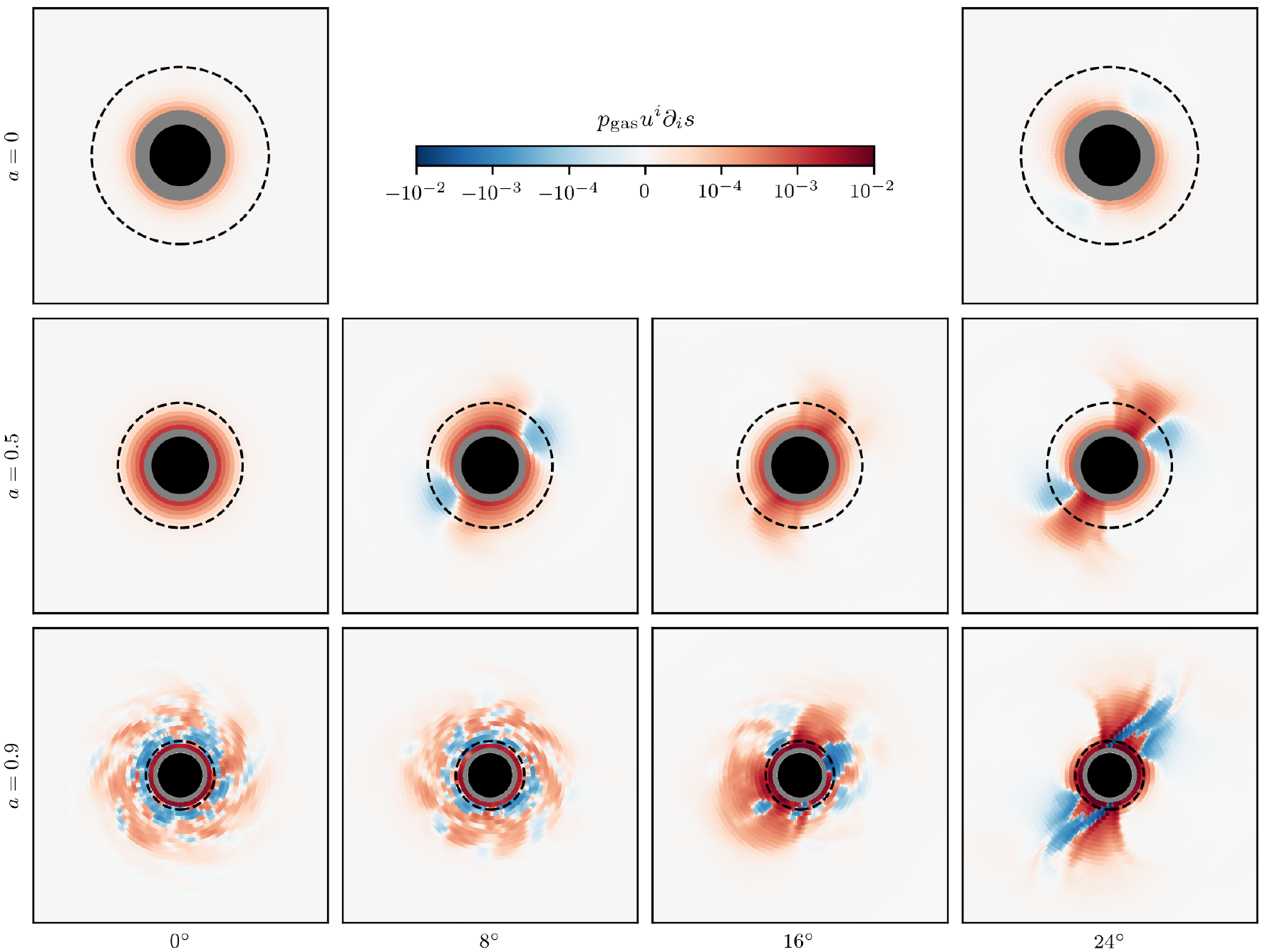}
  \caption{Irreversible heating rates, averaged within a scale height of the disk and plotted in the curved plane of the disk, for all the simulations. Each panel is $20$ gravitational radii wide. Each ring of a given radius is oriented with the angular momentum of that shell of material pointing out of the page and the ascending node placed directly up. The $m = 2$ pattern of positive entropy generation at high inclinations and nonzero spins indicates the location of the standing shocks. The concentric circles denote the event horizon, equatorial prograde photon sphere, and equatorial ISCO. \label{fig:q_src_images}}
\end{figure*}

Plots of the spatial distribution of this source term are shown in Figure~\ref{fig:q_src_images}. In order to make these plots, we orient each shell of material according to its own angular momentum. The longitude of ascending node at $\phi' = \pi/2$ is the upward vertical axis in the figures, with the angular momentum vector coming out of the page. As a result the plane of the figure is warped and twisted in $r,\theta,\phi$ coordinates. The source term is volume-averaged over a scale height in $\theta'$ at each point $r,\phi'$. The solid black circles indicate the event horizon, the gray shaded regions are inside the prograde photon sphere radius $\rphot$, and the dashed black lines denote the ISCO, all calculated in the black hole's midplane.

From the figure we can see a trend of stronger, more well-defined shocks at higher inclinations. Much of this happens outside the photon sphere, which can be used as a rough estimate for the boundary inside of which extra dissipation will have only negligible observable consequences \citep{Generozov2014}. In fact, the shock structure extends well beyond the ISCO.

\begin{figure*}
  \centering
  \includegraphics{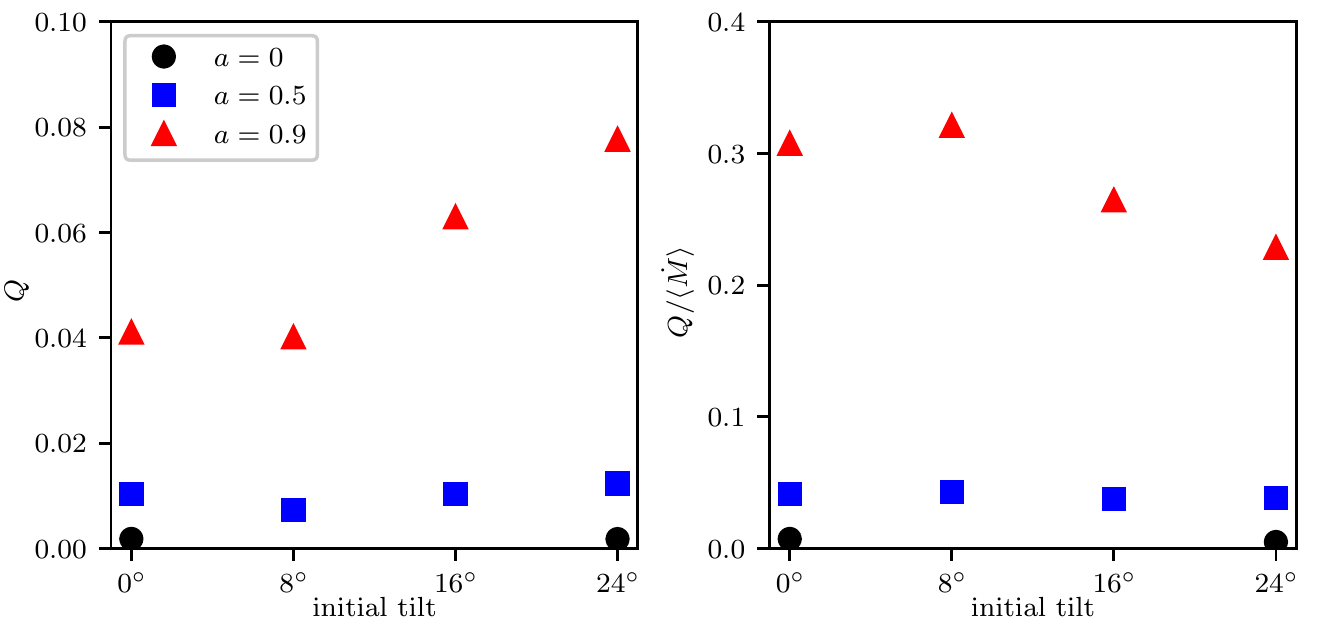}
  \caption{Left:\ integrated irreversible heating per unit coordinate time for the simulations. This is the integral of $-u_0 \pgas u^i \partial_i s$ from $\rphot$ to $r = 10$. For large, fixed spin, dissipation increases with increasing inclination, past about $8^\circ$. The two $a = 0$ points should physically be the same, and so their small difference ($2\%$) can be taken as an estimate of the numerical uncertainty of this measurement. Right:\ the same quantity, normalized by the time average of $\dot{M}$ over $3000 \leq t \leq 4000$ in the simulation. The decrease with tilt in the $a = 0.9$ case indicates the shocks enhance angular momentum transport even more than dissipation. \label{fig:e_src}}
\end{figure*}

More quantitatively, we can integrate the heat source term over the domain to gauge the extra heating caused by the shock. To do this, we must first incorporate our heating term into a proper tensor equation for energy evolution. In the fluid rest frame, we know $q$ must be the source for contravariant energy density $T^{00}$. As this source term should be the time component of a four-vector, it is more appropriate to say $q u^0$ is the source for $T^{00}$, as $u^0 = 1$ in this frame. More generally, the adiabatic evolution of conserved (covariant) energy-momentum density must then obey
\begin{equation}
  (\nabla_\mu \tensor{T}{^\mu_\nu})_\mathrm{adi} = -q u_\nu
\end{equation}
in all frames. Physically, the factor of $-u_0$ multiplying the right-hand side when considering energy ($\nu = 0$) accounts for gravitationally redshifting and Doppler shifting the released energy to a stationary observer at infinity. Thus we define the integrated source term
\begin{equation}
  Q \equiv \int -u_0 \pgas u^i (\partial_i s) \sqrt{-g} \, \dr\,\dth\,\dph,
\end{equation}
which agrees with \citet[\extref{45}]{Ressler2015}, where a fluid-frame heating rate is integrated to find a global quantity measured at infinity. We integrate from $\rphot$ to $r = 10$, with the upper bound chosen in order to focus on the shock in the region of the flow in steady state. $Q$ has units of energy per unit coordinate time. The results are plotted in the left panel of Figure~\ref{fig:e_src}. At low spins, the heating rate is not strongly affected by disk inclination, implying that whatever nonaxisymmetric shocks are created in these cases are too weak or too limited in extent to significantly alter the flow. For $a = 0.5$ the heating rate is $19\%$ larger at $24^\circ$ compared to $0^\circ$. However, the $a = 0.9$ cases show strong tilt dependence, with the $24^\circ$ heating rate $89\%$ larger than found at $0^\circ$. Thus there is significant extra heating due to this mechanism, but it only applies when spins and inclinations are sufficiently high. This agrees with expectations from \citet{Teixeira2014}, who found that no shocks form for $a = 0.1$.

The right panel normalizes $Q$ by the time average of $\dot{M}$ over the same time interval over which $Q$ was measured, so the plotted quantity has units of energy per unit mass. This provides a measure of the heating efficiency of the accretion flow (a proxy for the radiative efficiency). The result of increasing initial tilt is that both shock heating and accretion rate increase, and in fact that latter has a steeper dependence on tilt, resulting in \emph{lower} heating efficiencies. In going from $0^\circ$ to $24^\circ$, $Q / \ave{\dot{M}}$ decreases by $7\%$ for $a = 0.5$ and by $26\%$ for $a = 0.9$. At higher inclinations the shocks induce more angular momentum transport per unit dissipation. We note however that this trend only applies when considering heating at all latitudes. When dissipation is only measured near the midplane, it increases with inclination faster than $\dot{M}$. That is, the increased heating is more concentrated in the disk midplane than the increased mass flux.

\begin{figure*}
  \centering
  \includegraphics[width=0.45\textwidth]{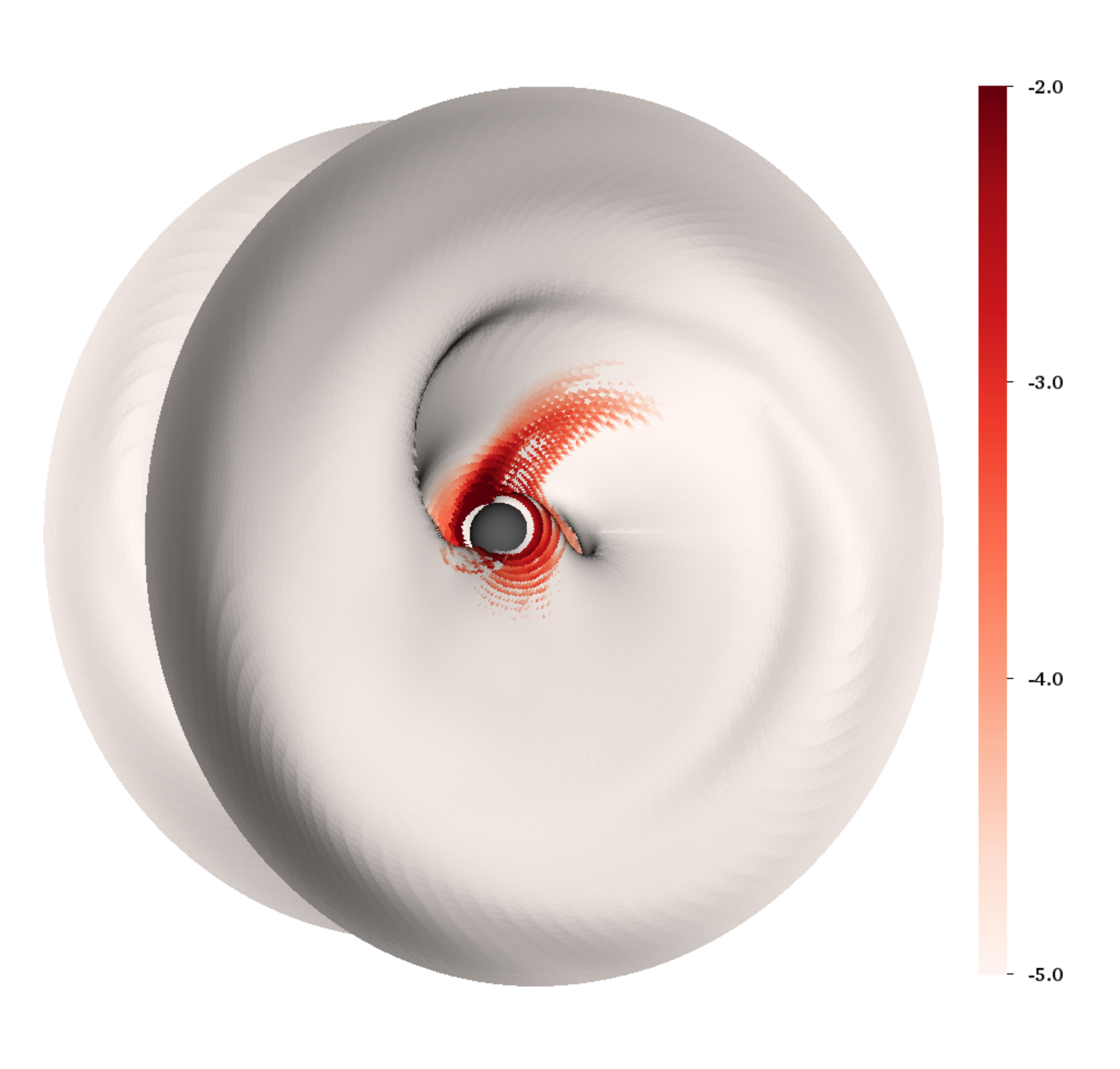}~
  \includegraphics[width=0.45\textwidth]{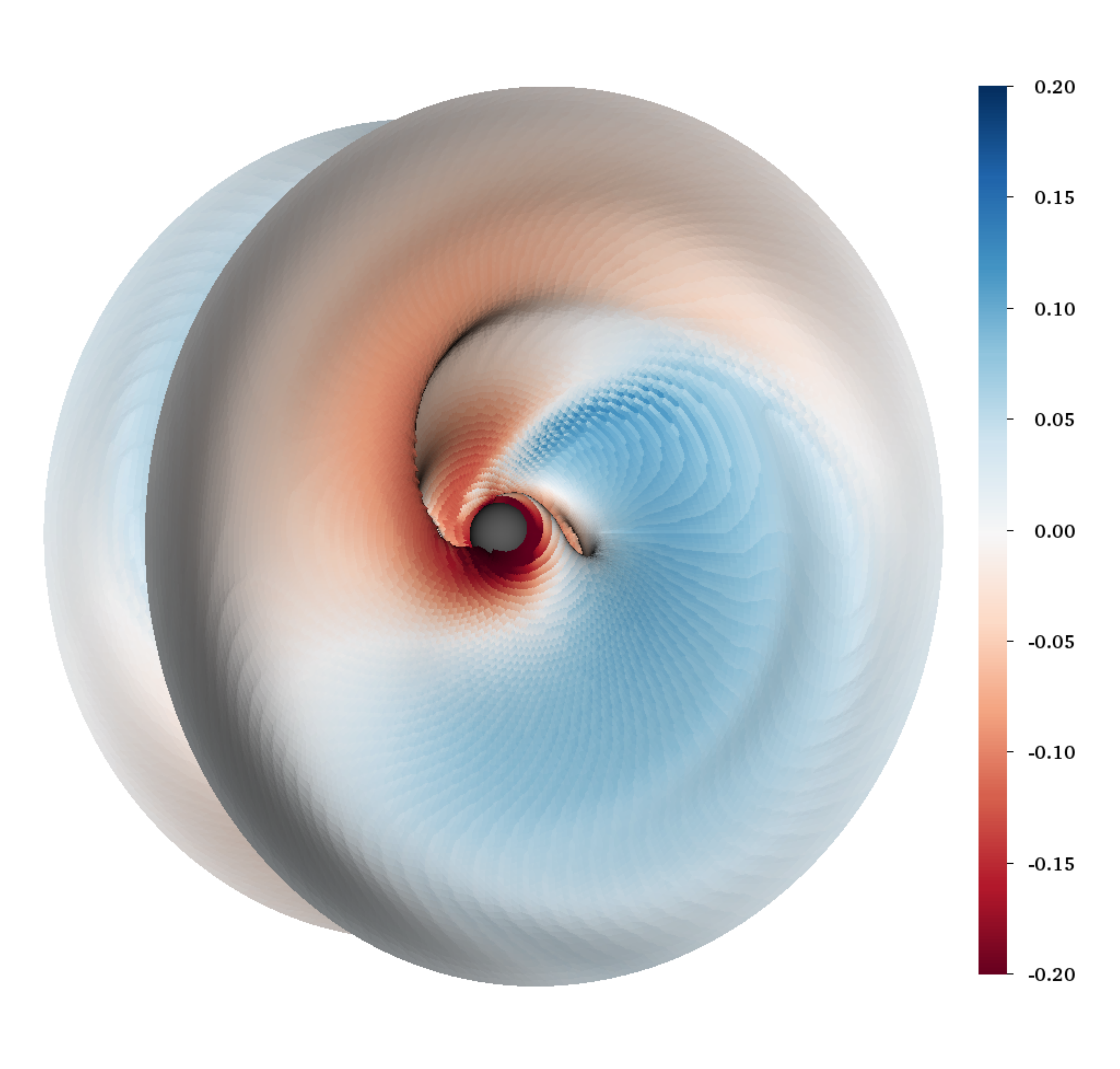}
  \caption{Three-dimensional visualizations of the $a = 0.9$, $24^\circ$ inclined simulation with the black hole spin out of the page and the ascending node of the disk at the top of each image. The colors show the local irreversible heating rate $\log_{10}(\pgas u^i \pp_i s)$ (left) and radial velocity $u^1$ (right) on surfaces of constant density $\rho = 0.05$ inside $r = 20$. The region of high entropy generation matches that of a rapid radial velocity change from outward to inward, occurring along the locus of apoapsides of eccentric orbits. The same pattern is repeated on the underside of the disk. \label{fig:3d}}
\end{figure*}

Figure~\ref{fig:3d} shows the three-dimensional structure of the shock in the $a = 0.9$, $24^\circ$ inclined case. Both panels are oriented looking down the black hole spin axis, with the disk ascending node vertical in the top half of the figure. The data is averaged over $3000 \leq t \leq 4000$. In both panels values are shown along the $\rho = 0.05$ isosurface within $r = 20$. On the left we show the heating term $\pgas u^i \pp_i s$ on a logarithmic scale, and on the right we show radial velocity $u^1$.

The radial velocity pattern indicates fluid elements are on eccentric orbits, in agreement with Figure~\ref{fig:vertical_u1} but more clearly seen here. This pattern is antisymmetric with respect to reflection through the midplane of the disk. At the disk antinodes, material that is at high latitudes with respect to the disk but near the black hole midplane is moving outward. When these fluid elements reach apoapsis near the disk's line of nodes, they encounter the shock and are abruptly redirected inward. At the same time we can consider material at the disk antinodes and above the disk but at very high black hole latitudes. This material is moving inward, but when it reaches periapsis near the disk line of nodes it changes radial direction smoothly with no shock. This picture of material shocking where eccentric orbits' apoapsides cluster (due to radially varying eccentricity) matches that of \citet{Fragile2008}.

Only one standing shock is visible in Figure~\ref{fig:3d}, with the other one hidden on the underside of the disk, $180^\circ$ away in azimuth. The shock locations correspond to where the material abruptly changes from moving outward to moving inward, indicating the shocks are significant causes of angular momentum transport in this simulation.

\begin{figure*}
  \centering
  \includegraphics{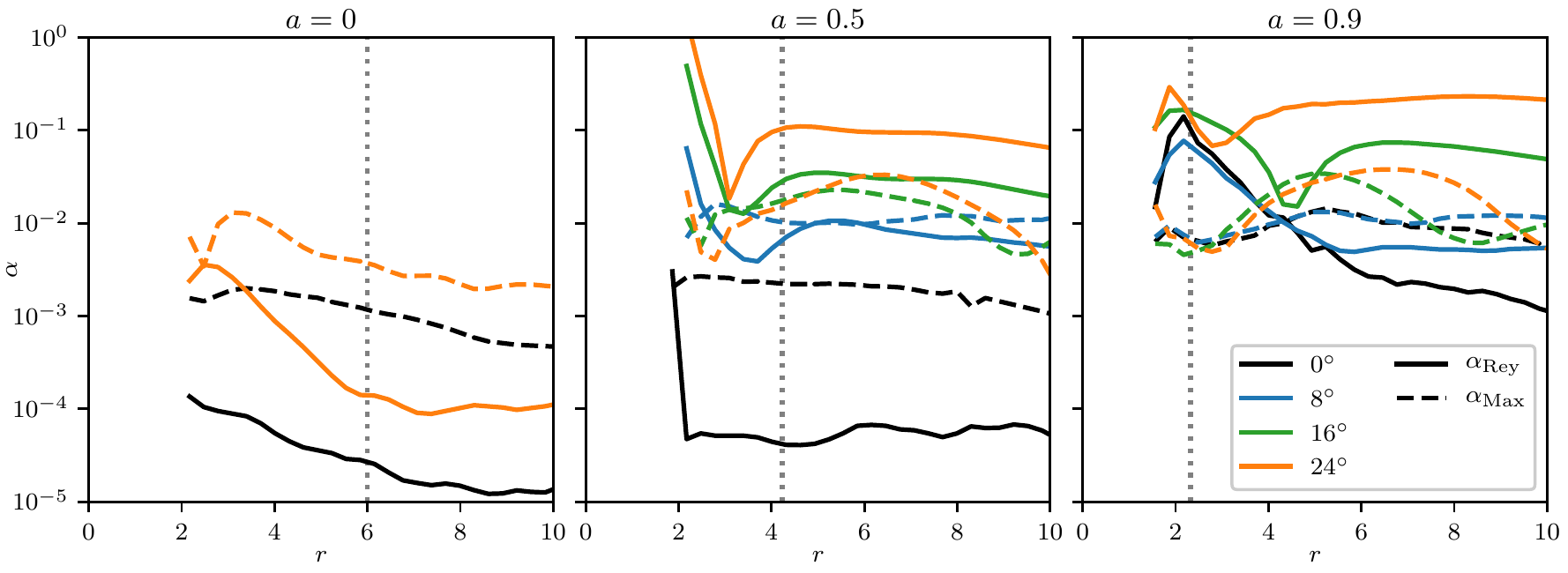}
  \caption{Effective accretion viscosities related to Reynolds and Maxwell stresses, time-averaged over $3000 \leq t \leq 4000$. As inclination angle and spin increase, Reynolds stresses become more important relative to Maxwell stresses. This is expected given the strengthening of the standing hydrodynamic shock. The vertical dotted lines indicate the radius of the equatorial ISCO. \label{fig:stress}}
\end{figure*}

The relative importance of the shocks can be examined by partitioning the stress-energy tensor into Reynolds and Maxwell components:
\begin{subequations} \begin{align}
  \tens{T}{_{\mathrm{Rey}}^{\mu\nu}} & \equiv \paren[\bigg]{\rho + \frac{\Gamma}{\Gamma-1} \pgas} \tens{u}{^\mu} \tens{u}{^\nu} + \pgas \tens{g}{^{\mu\nu}}, \\
  \tens{T}{_{\mathrm{Max}}^{\mu\nu}} & \equiv 2 \pmag \tens{u}{^\mu} \tens{u}{^\nu} + \pmag \tens{g}{^{\mu\nu}} - \tens{b}{^\mu} \tens{b}{^\nu}.
\end{align} \end{subequations}
We define $\alpharey$ and $\alphamax$ using the appropriate stress-energy terms in the definition
\begin{equation}
  \alpha \equiv \frac{\int \abs{T^{x''y''}} / (\pgas + \pmag) \times \rho \sqrt{-g} \, \dth\,\dph}{\int \rho \sqrt{-g} \, \dth\,\dph}.
\end{equation}
Here the tensor is evaluated in a Cartesian frame comoving with the average fluid as described in the appendix, \S\ref{sec:alpha}. The coordinates are oriented such that $z''$ points along the angular momentum direction at that radius and $y''$ points along the ascending node, and so this definition (corresponding to $\alphaacc$ in the appendix) captures the radial transport of angular momentum that is aligned with the disk axis and drives accretion. These values are plotted in Figure~\ref{fig:stress}. In the nontrivially tilted cases Reynolds stress is generally larger than Maxwell stress. This difference is especially striking in the innermost parts of the $a = 0.9$ tilted simulations, where the shock is strongest.

% Discussion
\section{Discussion}
\label{sec:discussion}

We have run ten simulations showing how the same initial magnetized fluid evolves as a function of black hole spin and disk inclination. In all cases with nontrivial inclinations, the disks evolve to a highly warped and twisted steady state that precesses rigidly. While it is clear that torques are being transmitted through the disks, redistributing angular momentum, the shapes are not readily described by a straightforward application of linear bending wave theory. Given that we did not go to the most extreme spins, nor to very high inclinations, we expect many naturally occurring systems to challenge such a theory with their nonlinear behavior.

For these inefficiently cooled, thick accretion flows, there is no observable Bardeen--Petterson effect, nor do the disks break. Rather, the inner parts of the disk smoothly become more inclined than the outer parts (Figure~\ref{fig:vertical_rho}). This can be explained by angular momentum transport not being evenly distributed in azimuth, as happens when standing shocks form at the line of nodes, combined with the peculiar shape of inclined circular geodesics in Kerr spacetime. Analytic models that assume azimuthally symmetric coupling between adjacent rings of material can miss this effect.

All of our simulations have a low-density polar region. When there is no spin the relatively little amount of net vertical magnetic flux means this region consists of material falling inward, though an outflow turns on with even moderate spins. At $a = 0.5$, this moderate outflow can be entirely disrupted by any inclination in the disk; tilted disks can suppress otherwise moderate outflows. With $a = 0.9$ the outflow is strong enough to be formed even in the presence of a tilt. These outflows align with the disk rather than the spin of the black hole, in agreement with \citet{Liska2018}. However, the large-scale, average magnetic field in some cases evolves from being initially aligned with the disk to partially aligning with the black hole. Further investigation with simulations that achieve high resolution in the polar regions can help quantify to what extent the velocity and magnetic structure of the outflow depend on the tilt of the disk.

Our results confirm the presence of standing shocks, as found in \citet{Fragile2008} and later works. These shocks considerably increase the heating of material and the hydrodynamic angular momentum transport in the inner $5\text{--}10$ gravitational radii of the disk, but only at sufficiently high spin and only when the inclination is larger than about $8^\circ$. The effects are more pronounced at higher inclination angles, and they can be the dominant driving mechanism for both dissipation and accretion.

Importantly, these shocks can be better at redirecting fluid elements than stopping them. This can lead to more angular momentum transport and thus accretion per unit dissipation, mimicking the lower heating efficiencies of systems with lower black hole spins. As noted above, however, this trend reverses if the higher latitudes of these accretion flows are neglected. It is possible that a given heating efficiency can correspond to a range of black hole spins. While tilted disk simulations with in-situ radiation will be required to verify these trends, but it is clear that one should be cautious when interpreting possibly tilted flows using untilted models.

In the limit where accretion is dominated by these shocks ($a \gtrsim 0.9$ and $i \gtrsim 16^\circ$), it is possible that simple analytic models can be developed to capture the correct behavior. That is, there may be regimes where the complexity of MRI turbulence can be neglected, not because it can be replaced by an isotropic viscosity, but because in reality transport becomes dominated by a mechanism more amenable to modeling. Other processes like the MRI are, however, still needed to drive material close to the black hole in the first place.

Given the presence of standing shocks, we expect to see time variability in accretion, based on the results of \citet{Henisey2012}. Here we have focused on large-scale properties that stay constant in time, leaving the examination of the variability present in these simulations for a future work. We also suggest that a parameter survey similar to this be done with more efficiently cooled disks. This can help address whether the lack of disk breaking we see here, where GR and MHD processes are being modeled accurately, extends to thinner disks.

In summary, the magnitude of tilt does not strongly affect dimensionless warping and twisting, nor does it qualitatively change the behavior of inclination and precession angles. It affects the velocity of polar outflows, while having less of an effect on their magnetization. Most importantly, large tilts can increase the efficacy of standing shocks in heating the material and transporting angular momentum.

% Acknowledgments
\acknowledgments

We thank S.~M. Ressler for pointing out the subtleties of entropy and heating in a GR context. This research was supported in part by the National Science Foundation under grants NSF~PHY-1748958, NSF~AST~13-33612, and NSF~AST~1715054; by \emph{Chandra} theory grant TM7-18006X from the Smithsonian Institution; and by a Simons Investigator award from the Simons Foundation (EQ). This work used the Extreme Science and Engineering Discovery Environment (XSEDE) Comet at the San Diego Supercomputer Center through allocation AST170012.

% Software
\software{\athena{} \citep{White2016}}

% Appendices
\appendix

% Nodal precession of geodesics in Kerr spacetime
\section{Nodal Precession of Geodesics in Kerr Spacetime}
\label{sec:nodal}

Here we outline the numerical evaluation of the nodal precession rate in Kerr spacetime. The procedure here is exact, as opposed to the well-known Lense--Thirring formula, which is formulated in the linear regime and breaks down for large $a$ and small $r$.

We will work in Boyer--Lindquist coordinates until the end of this section. In this system, the geodesics of massive particles obey the well-known equations \citep[cf.][\extref{1--3}]{Wilkins1972}
\begin{subequations} \label{eq:motion} \begin{align}
  \label{eq:motion:0} \Sigma u^0 & = a \paren[\big]{\ell - a e \sin^2\!\theta} + \frac{1}{\Delta} \paren[\big]{r^2 + a^2} P, \\
  \label{eq:motion:1} \Sigma u^1 & = \pm\sqrt{R}, \\
  \label{eq:motion:2} \Sigma u^2 & = \pm\sqrt{\Theta}, \\
  \label{eq:motion:3} \Sigma u^3 & = \frac{\ell}{\sin^2\!\theta} - a e + \frac{a}{\Delta} P,
\end{align} \end{subequations}
where we define
\begin{subequations} \label{eq:motion_def} \begin{align}
  \label{eq:motion_def:delta} \Delta & \equiv r^2 - 2 M r + a^2, \\
  \label{eq:motion_def:sigma} \Sigma & \equiv r^2 + a^2 \cos^2\!\theta, \\
  \label{eq:motion_def:p} P & \equiv e \paren[\big]{r^2 + a^2} - \ell a, \\
  \label{eq:motion_def:r} R & \equiv P^2 - \Delta \paren[\Big]{r^2 + q + \paren[\big]{\ell - a e}^2}, \\
  \label{eq:motion_def:theta} \Theta & \equiv q - \paren[\bigg]{a^2 \paren[\big]{1 - e^2} + \frac{\ell^2}{\sin^2\!\theta}} \cos^2\!\theta,
\end{align} \end{subequations}
and $e$, $l$, and $q$ are constants of the motion.

Consider a particle initially at position $x_{(0)}^\mu = (0, r, \pi/2 + i, 0)$, for fixed radius $r$ and inclination $i$, and with velocity $u_{(0)}^\mu = (u_{(0)}^0, 0, 0, u_{(0)}^3)$. We want the orbit to stay at the same radius for all time, and so the radial geodesic equation becomes
\begin{equation}
  \tensor*{\Gamma}{^1_{\mu\nu}} \tens{u}{_{(0)}^\mu} \tens{u}{_{(0)}^\nu} = 0,
\end{equation}
which is a quadratic equation for $u_{(0)}^3 / u_{(0)}^0$ (taking the positive root for prograde motion). This can be used with the velocity normalization
\begin{equation}
  \tens{g}{_{\mu\nu}} \tens{u}{_{(0)}^\mu} \tens{u}{_{(0)}^\nu} = -1,
\end{equation}
to find $u_{(0)}^0$ and $u_{(0)}^3$ independently.

Equations~\eqref{eq:motion:0} and~\eqref{eq:motion:3} can be rearranged, together with \eqref{eq:motion_def:delta} and \eqref{eq:motion_def:p}, to yield
\begin{subequations} \begin{align}
  \paren[\bigg]{\frac{1}{\Delta} \paren[\big]{r^2 + a^2}^2 - a^2 \sin^2\!\theta} e - \frac{2Ma r}{\Delta} \ell & = \Sigma \tens{u}{_{(0)}^0}, \\
  \frac{2Ma r}{\Delta} e + \paren[\bigg]{\frac{1}{\sin^2\!\theta} - \frac{a^2}{\Delta}} \ell & = \Sigma \tens{u}{_{(0)}^3}. \\
\end{align} \end{subequations}
This linear system can easily be solved to find the specific energy and angular momentum constants $e$ and $\ell$. The Carter constant $q$ then follows from \eqref{eq:motion_def:theta} and knowing $\Theta = 0$ from \eqref{eq:motion:2} (or from \eqref{eq:motion_def:p} and \eqref{eq:motion_def:r} knowing $R = 0$ from \eqref{eq:motion:1}).

With the constants of motion in hand, we can compute the relative rate of change between the two angular coordinates:
\begin{equation}
  \frac{\dph}{\dth} \equiv \frac{u^3}{u^2} = \pm\paren[\bigg]{\frac{\ell}{\sin^2\!\theta} - a e + \frac{a e (r^2 + a^2) - \ell a^2}{r^2 - 2 M r + a^2}} \paren[\bigg]{q - \paren[\bigg]{a^2 \paren[\big]{1 - e^2} + \frac{\ell^2}{\sin^2\!\theta}} \cos^2\!\theta}^{-1/2}.
\end{equation}
This formula is similar to \extref{25} from \citet{Wilkins1972}, but we include the spin dependence rather than assume an extremal black hole. The overall negative sign should be taken for prograde motion with $i > 0$, as the particle will initially start to move toward smaller $\theta$ as it moves forward in $\phi$. By symmetry we know $\phi$ advances by the same amount in the first quarter of the vertical oscillation as in all other quarters, and so the total azimuthal precession in one vertical period is
\begin{equation}
  \Delta\phi = 4 \int_{\pi/2+i}^{\pi/2} \frac{\dph}{\dth} \, \dth - 2\pi.
\end{equation}
The corresponding change in proper time for the particle is similarly calculated:
\begin{equation}
  \Delta\tau = 4 \int_{\pi/2+i}^{\pi/2} \frac{1}{u^2} \, \dth.
\end{equation}

Dividing $\Delta\phi$ by $\Delta\tau$ gives us $\dph/\dtau$, the average nodal precession of particles at a given radius and inclination around a black hole with a given spin. This is easily converted to Kerr--Schild coordinates via
\begin{equation}
  \frac{\dph_\mathrm{KS}}{\dtau} = \frac{\dph_\mathrm{BL}}{\dtau} + \frac{a}{\Delta} u^1.
\end{equation}
In particular, we can neglect the difference between Kerr--Schild and Boyer--Lindquist coordinates to the extent our approximation of vanishing radial velocity is appropriate.

For comparison, the Lense--Thirring result is a gravitomagnetic nodal advance of $4\pi a r^{-3/2}$ per orbit. Alternatively, a more accurate estimate of $2\pi (1 - \Omega_\theta / \Omega)$ can be used (see \S\ref{sec:frequencies} for the definitions of $\Omega$ and $\Omega_\theta$), taking into account the retrograde contribution from the black hole's quadrupole moment. Around a black hole of spin $a = 0.9$ at a radius of $r = 5$, these formulas predict $\Delta\phi \approx 58.0^\circ$ and $\Delta\phi \approx 43.0^\circ$, while the exact result ranges from approximately $49.0^\circ$ to $49.9^\circ$ as the inclination ranges from $8^\circ$ to $24^\circ$.

% Dynamical frequencies for circular geodesics in Kerr spacetime
\section{Dynamical Frequencies for Circular Geodesics in Kerr Spacetime}
\label{sec:frequencies}

For reference we give the dynamical frequencies for prograde circular equatorial orbits in Kerr spacetime. All formulas here are the same in both Kerr--Schild and Boyer--Lindquist coordinates, since the two $3{+}1$ metrics induce the same $2{+}1$ metric on any given hypersurface of constant $r$.

The orbital, epicyclic, and vertical frequencies with respect to coordinate time are given by \citet{Okazaki1987} and \citet{Kato1990}:
\begin{subequations} \begin{align}
  \Omega & \equiv \frac{\dph}{\dt} = \paren[\Bigg]{\sqrt{\frac{r^3}{M}} + a}^{-1}, \\
  \Omega_r & = \Omega \paren[\Bigg]{1 - 6 \paren[\bigg]{\frac{r}{M}}^{-1} + \frac{8a}{M} \paren[\bigg]{\frac{r}{M}}^{-3/2} - \frac{3a^2}{M^2} \paren[\bigg]{\frac{r}{M}}^{-2}}^{1/2}, \\
  \Omega_\theta & = \Omega \paren[\Bigg]{1 - \frac{4a}{M} \paren[\bigg]{\frac{r}{M}}^{-3/2} + \frac{3a^2}{M^2} \paren[\bigg]{\frac{r}{M}}^{-2}}^{1/2}.
\end{align} \end{subequations}
We can convert these to frequencies with respect to an orbiting particle's proper time via the Lorentz factor $\gamma \equiv \dt/\dtau$. As the particle's velocity is $u^\mu = (\gamma, 0, 0, \gamma \Omega)$, the normalization $g_{\mu\nu} u^\mu u^\nu = -1$ yields
\begin{equation}
  \gamma = \paren[\big]{-\tens{g}{_{00}} - 2 \tens{g}{_{03}} \Omega - \tens{g}{_{33}} \Omega^2}^{-1/2}.
\end{equation}
With $\gamma$ we can convert $\Omega$ to the orbital frequency per unit proper time $\omega \equiv \dph/\dtau$, and likewise for the other frequencies:
\begin{subequations} \begin{align}
    \omega & = \gamma \Omega, \\
    \omega_r & = \gamma \Omega_r, \\
    \omega_\theta & = \gamma \Omega_\theta.
\end{align} \end{subequations}

\begin{figure*}
  \centering
  \includegraphics{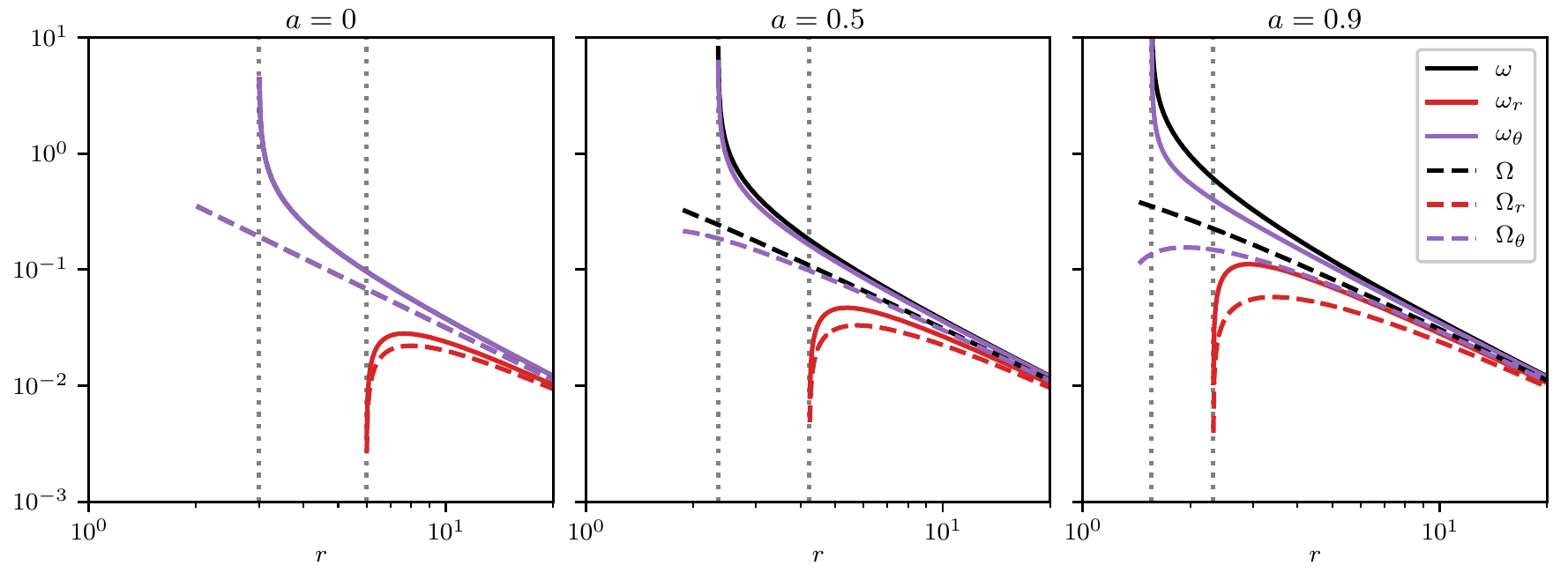}
  \caption{Orbital, epicyclic, and vertical frequencies as functions of radius. The vertical dotted lines denote the prograde photon orbit at smaller $r$ and the equatorial ISCO at larger $r$. In the left panel the vertical frequencies are exactly the same as the orbital frequencies. \label{fig:frequencies}}
\end{figure*}

Plots of these frequencies for the spins considered in our simulations are shown in Figure~\ref{fig:frequencies}. These enter into the linear warp theory as deviations of epicyclic and vertical frequencies from the orbital frequency. Since the epicyclic frequency necessarily diverges at the ISCO, its deviations are particularly large for the inner parts of the disk, to the point where linear theory may no longer be applicable even for very small inclinations.

% Definition of effective viscosity
\section{Definition of Effective Viscosity}
\label{sec:alpha}

In order to define an $\alpha$-viscosity in a way appropriate for tilted disks and GR, we adopt and modify a procedure described in \citet{Penna2013}.

We begin with the fluid velocity in Kerr--Schild coordinates $u_\mathrm{KS}^\mu$. These are easily transformed to Boyer--Lindquist coordinates via
\begin{subequations} \begin{align}
  \ubla & = \uksa - \frac{2 M r}{r^2 - 2 M r + a^2} \uksb, \\
  \ublb & = \uksb, \\
  \ublc & = \uksc, \\
  \ubld & = \uksd - \frac{a}{r^2 - 2 M r + a^2} \uksb.
\end{align} \end{subequations}
Knowing the Euler angles $i$ and $\phi_0$ at each radius (see \S\ref{sec:measuring}), we can convert to the disk-aligned frame. Specifically, we need the components
\begin{subequations} \begin{align}
  \ublbp & = \ublb, \\
  \ubldp & = \frac{\pp\phi'}{\pp\theta} \ublc + \frac{\pp\phi'}{\pp\phi} \ubld.
\end{align} \end{subequations}
For completeness, we give all the components of the Jacobian corresponding to \eqref{eq:spherical_transformation}:
\begin{subequations} \begin{align}
  \sin\theta' \frac{\pp\theta'}{\pp\theta} & = -\sin i \cos\phi_0 \cos\theta \cos\phi - \sin i \sin\phi_0 \cos\theta \sin\phi + \cos i \sin\theta, \\
  \sin\theta' \frac{\pp\theta'}{\pp\phi} & = -\sin i \sin\phi_0 \sin\theta \cos\phi + \sin i \cos\phi_0 \sin\theta \sin\phi, \\
  \sin^2\!\theta' \frac{\pp\phi'}{\pp\theta} & = \sin i \sin\phi_0 \cos\phi - \sin i \cos\phi_0 \sin\phi, \\
  \sin^2\!\theta' \frac{\pp\phi'}{\pp\phi} & = -\sin i \cos\phi_0 \cos\theta \sin\theta \cos\phi - \sin i \sin\phi_0 \cos\theta \sin\theta \sin\phi + \cos i \sin^2\!\theta, \\
  \sin\theta \frac{\pp\theta}{\pp\theta'} & = \sin i \cos\theta' \cos\phi' + \cos i \sin\theta', \\
  \sin\theta \frac{\pp\theta}{\pp\phi'} & = -\sin i \sin\theta' \sin\phi', \\
  \sin^2\!\theta \frac{\pp\phi}{\pp\theta'} & = \sin i \sin\phi', \\
  \sin^2\!\theta \frac{\pp\phi}{\pp\phi'} & = \sin i \cos\theta' \sin\theta' \cos\phi' + \cos i \sin^2\!\theta'.
\end{align} \end{subequations}

We next define the mean velocity components
\begin{subequations} \begin{align}
  \ubblbp & \equiv \frac{1}{2\pi} \int_0^{2\pi} \ublbp \, \dph', \\
  \ubblcp & \equiv 0, \\
  \ubbldp & \equiv \frac{1}{2\pi} \int_0^{2\pi} \ubldp \, \dph'.
\end{align} \end{subequations}
This averaging only yields meaningful results with tilted components and along rings that vary only in the coordinate $\phi'$. We then find the corresponding time components $\bar{u}_\mathrm{BL}^{0'}$ via
\begin{equation}
  \tens{g}{^{\mathrm{BL}}_{\mu'\nu'}} \ubblmp \ubblnp = -1.
\end{equation}
Note that while the radial and azimuthal components of the mean velocity vary with just $r$ and $\theta'$, the time component will in general vary with $\phi'$ as well. These four components are then transformed back into the untilted coordinates:
\begin{subequations} \begin{align}
  \ubbla & = \ubblap, \\
  \ubblb & = \ubblbp, \\
  \ubblc & = \frac{\pp\theta}{\pp\theta'} \ubblcp + \frac{\pp\theta}{\pp\phi'} \ubbldp, \\
  \ubbld & = \frac{\pp\phi}{\pp\theta'} \ubblcp + \frac{\pp\phi}{\pp\phi'} \ubbldp.
\end{align} \end{subequations}

We can now construct the transformation from untilted Boyer--Lindquist coordinates to an orthonormal frame following an appropriately defined mean flow, as was first done in \citet{Krolik2005}. (There is still a degree of arbitrariness in rotating the radial and poloidal directions of this frame, as has been noted by previous authors.) We present the formulas in notation more closely following \citet{Penna2013}, correcting one of the normalizations. The transformation components are
\begin{subequations} \label{eq:fluid_transformation} \begin{align}
  \tens{e}{_{0''}^\mu} & = \ubblm, \\
  \tens{e}{_{1''}^\mu} & = \frac{s}{N_1} \paren[\Big]{\ubblbl \ubbla, 1 + \ubblbl \ubblb + \ubblcl \ubblc, 0, \ubblbl \ubbld}, \\
  \tens{e}{_{2''}^\mu} & = \frac{1}{N_2} \paren[\Big]{\ubblcl \ubbla, \ubblcl \ubblb, 1 + \ubblcl \ubblc, \ubblcl \ubbld}, \\
  \tens{e}{_{3''}^\mu} & = \frac{1}{N_3} \paren[\Big]{-\ubbldl / \ubblal, 0, 0, 1},
\end{align} \end{subequations}
with the definitions
\begin{subequations} \label{eq:fluid_normalization} \begin{align}
  s & = -\sgn\paren[\Big]{\ubblal \ubbla + \ubbldl \ubbld}, \\
  N_1 & = \paren[\bigg]{\ubblbl \ubblbl \paren[\Big]{\ubblal \ubbla + \ubbldl \ubbld} + \tens{g}{_{11}} \paren[\Big]{\ubblal \ubbla + \ubbldl \ubbld}^2}^{1/2}, \\
  N_2 & = \paren[\bigg]{g_{22} \paren[\Big]{1 + \ubblcl \ubblc}}^{1/2}, \\
  N_3 & = \paren[\bigg]{g_{00} \paren[\Big]{\ubbldl / \ubblal}^2 - 2 \tens{g}{_{03}} \ubbldl / \ubblal + \tens{g}{_{33}}}^{1/2}.
\end{align} \end{subequations}
The velocity at any point can be transformed into this fluid frame according to
\begin{equation}
  \tens{u}{^{i''}} = \tens{g}{^{\mathrm{BL}}_{\mu\nu}} \tens{e}{_{i''}^\mu} \ubln.
\end{equation}
The magnetic field components $b^{i''}$ are found in the same way.

We perform one last transformation to Cartesian coordinates:
\begin{subequations} \begin{align}
  u^{x''} & = \sin\theta' \cos\phi' u^{1''} + \cos\theta' \cos\phi' u^{2''} - \sin\phi' u^{3''}, \\
  u^{y''} & = \sin\theta' \sin\phi' u^{1''} + \cos\theta' \sin\phi' u^{2''} + \cos\phi' u^{3''}, \\
  u^{z''} & = \cos\theta' u^{1''} - \sin\theta' u^{2''}
\end{align} \end{subequations}
(and similarly for the magnetic field). Here the $z''$-direction, like the $z'$-direction, points in the direction of the disk angular momentum. Thus $x''y''$-stress corresponds to radial transport of $z''$-angular momentum, driving accretion. Similarly the $y''$-direction points to the ascending node, so $x''z''$-stress governs rotation about this axis, aligning the disk with the black hole. Finally, $y''z''$-stress corresponds to precessing the disk angular momentum about the black hole axis. These considerations justify the definitions
\begin{subequations} \begin{align}
  \alphaacc & \equiv \frac{\int T^{x''y''} / (\pgas + \pmag) \times \rho \sqrt{-g} \, \dth\,\dph}{\int \rho \sqrt{-g} \, \dth\,\dph}, \\
  \alphapre & \equiv \frac{\int T^{x''z''} / (\pgas + \pmag) \times \rho \sqrt{-g} \, \dth\,\dph}{\int \rho \sqrt{-g} \, \dth\,\dph}, \\
  \alphaali & \equiv \frac{\int T^{y''z''} / (\pgas + \pmag) \times \rho \sqrt{-g} \, \dth\,\dph}{\int \rho \sqrt{-g} \, \dth\,\dph},
\end{align} \end{subequations}
for the accretion, precession, and alignment effective viscosities. Though the calculation of these relativistic effective viscosities are rather involved, they still broadly agree with the Newtonian equivalents. For example, $\alphapre$ is small in magnitude and suffers large cancellations when averaging, as was found with $\alpha_*$ derived from $r\theta$-stresses in \citet{Sorathia2013}.

% References
\bibliographystyle{aasjournal}
\bibliography{references}

\end{document}